\documentclass{article}

\usepackage[dvipdfmx]{graphicx}
\usepackage{amsmath}        
\usepackage{amssymb}        
\usepackage{cite}
\usepackage[dvips]{color}
\usepackage{fancyhdr}

\def\erase#1{{}}
\def\EqArrerase#1{{}}
\def\correct#1#2{#2}
\def\add#1{#1}

\makeatletter
 
 \@addtoreset{equation}{section}
\makeatother

\def\B{{\rm B}}
\def\T{{\rm T}}

\def\nn{\nonumber\\}

\def\calM{{\cal M}}
\def\calK{{\cal K}}
\def\calG{{\cal G}}
\def\calP{{\cal P}}

\def\calJ{{\cal J}}
\def\calT{{\cal T}}
\def\calE{{\cal E}}

\def\calF{{\cal F}}
\def\QB{Q_{\rm B}}
\def\Brs{{\boldsymbol\delta}_{\rm B}}
\def\Trs{{\boldsymbol\delta}}
\def\BrsN{{\boldsymbol\delta}_{\rm N}}
\def\GF{{\rm GF}}
\def\FP{{\rm FP}}
\def\GFP{{\rm GF+FP}}

\def\GL{{G\kern-.12em L\kern-.04em}}
\def\OSp{{O\kern-.11em S\kern-.04em p}}
\def\IOSp{{I\kern-.06em O\kern-.11em S\kern-.04em p}}
\def\MN{{M\kern-.14em N}}
\def\NM{{N\kern-.14em M}}
\def\NL{{N\kern-.14em L}}
\def\LN{{L\kern-.11em N}}
\def\ML{{M\kern-.14em L}}
\def\LM{{L\kern-.11em M}}
\def\RN{{R\kern-.11em N}}
\def\NR{{N\kern-.14em R}}
\def\RM{{R\kern-.11em M}}
\def\MR{{M\kern-.14em R}}
\def\RL{{R\kern-.11em L}}
\def\LR{{L\kern-.11em R}}
\def\RS{{R\kern-.11em S}}
\def\SR{{S\kern-.11em R}}
\def\SN{{S\kern-.11em N}}
\def\NS{{N\kern-.11em S}}
\def\SM{{S\kern-.11em M}}
\def\MS{{M\kern-.11em S}}
\def\SL{{S\kern-.11em L}}
\def\LS{{L\kern-.11em S}}
\def\sqr#1#2{{\vcenter{\hrule height.#2pt
      \hbox{\vrule width.#2pt height#1pt \kern#1pt
          \vrule width.#2pt}
      \hrule height.#2pt}}}
\def\bra0{\langle0|}
\def\ket0{|0\rangle}
\def\soeji#1_#2#3{#1_{#2}\cdots#1_{#3}}
\def\longgLRarrow{\longleftarrow\kern-3pt\relbar\kern-3pt\relbar\kern-3pt%
\longrightarrow}
\def\longLRarrow{\longleftarrow\kern-3pt\relbar\kern-3pt\longrightarrow}
\def\longLarrow{\longleftarrow\kern-3pt\relbar\kern-3pt\relbar\kern-3pt\relbar}
\def\longRarrow{\relbar\kern-3pt\relbar\kern-3pt\relbar\kern-3pt\longrightarrow}
\def\bothDer#1#2#3{%
\overset{\kern-.7em\stackrel{#1}{#2}}{\partial_{#3}}}
 
\makeatletter
 
 \@addtoreset{equation}{section}
\makeatother

\begin{document}
\thispagestyle{fancy}

\title{Noether Currents and Maxwell-type Equations of Motion in Higher Derivative Gravity Theories}

\author{Taichiro Kugo
\footnote{Electronic address: kugo@yukawa.kyoto-u.ac.jp}
\\
{\it\small
\begin{tabular}{c}
Yukawa Institute for Theoretical Physics, Kyoto University, Kyoto 606-8502, Japan
\end{tabular}
}
}

\maketitle

\thispagestyle{fancy}

\begin{abstract}
In general coordinate invariant gravity theories whose Lagrangians contain 
arbitrarily high order derivative fields, 
the Noether currents for the global translation and for the Nakanishi's 
$\OSp(8|8)$ choral symmetry containing the BRS symmetry as its member, are 
constructed. We generally show that for each of those Noether currents 
a suitable linear combination of equations of motion can be brought into 
the form of Maxwell-type field equation possessing the Noether current 
as its source term. 
\end{abstract}

\section{Introduction}

The equation of motion for the Yang-Mills field $A^a_\mu$ 
in the covariant gauge is 
given in the form
\begin{eqnarray}
\EqArrerase{D_\nu F^{a\mu\nu} + \partial_\mu B^a -ig f_{abc}\partial_\mu\bar c^b\cdot c^c = gj^a_\mu\,,} 
\add{D^\nu F^{a}_{\ \mu\nu} + \partial_\mu B^a -ig f^a{}_{bc}\partial_\mu\bar c^b\cdot c^c = gj^a_\mu\,,}
\end{eqnarray}
where 
\correct{$D_\nu F^{a\mu\nu}$}{$D^\nu F^a_{\ \mu\nu}$} is the covariant divergence of the field strength 
\correct{$F^{a\mu\nu}$}{$F^a_{\ \mu\nu}$},  
$j^a_\mu$ is the color current from the matter field, and 
$B^a, c^a$ and $\bar c^a$ are Nakanishi-Lautrup (NL), Faddeev-Popov (FP) 
ghost and anti-ghost fields, respectively. This equation was first noted by 
Ojima\cite{Ojima:1978hy} to be rewritten into the form of Maxwell-type equation of motion:
\begin{eqnarray}
\EqArrerase{
\partial_\nu F^{a\,\mu\nu} + \{ \QB, D_\mu\bar c^a \}  = gJ^a_\mu\,.} 
\add{
\partial^\nu F^a{}_{\mu\nu} + \{ \QB, D_\mu\bar c^a \}  = gJ^a_\mu\,.} 
\label{eq:Maxwell-typeYMeom}
\end{eqnarray}
Here, $\QB$ is the BRS charge and $J^a_\mu$ in the RHS is the Noether current 
for the global gauge transformation 
($=$ color rotation) under which all the gauge field $A_\mu^a$, NL and 
FP ghost fields, $B^a, c^a, \bar c^a$, transform as adjoint representations, 
given by
\begin{equation}
J^a_\mu = (A^\nu\times F_{\nu\mu})^a +j^a_\mu+(A_\mu\times B)^a -i(\bar c\times D_\mu c)^a 
+i(\partial_\mu\bar c\times c)^a\,,
\end{equation}
with 
\correct{$(A\times B)^a \equiv f_{abc}A^bB^c$} 
{$(A\times B)^a \equiv f^a{}_{bc}A^bB^c$}. 
This form of YM field equation (\ref{eq:Maxwell-typeYMeom}) is particular, 
firstly in the simple divergence form for the field strength, 
$\partial_\nu F^{a\,\mu\nu}$, and secondly 
in the BRS exact form for the NL and FP ghost contribution terms. 

This form of YM field equation, which we call ``Maxwell-type equation of 
motion", played very important rolls in discussing\cite{Kugo:1978wp,Kugo:1979gm}
\begin{enumerate}
\item existence of {\it elementary BRS quartet} of asymptotic fields
\item spontaneous breaking of color symmetry and Higgs phenomenon 
\item unbroken color symmetry and color confinement.
\end{enumerate} 
From the technical viewpoint also, it was useful to simplify the computations 
of equal time commutation (ETC) relations for some field variables as well as 
to derive Ward-Takahashi identities.

Also in gravity theory, there is a beautiful canonical formulation given by 
Nakanishi in a series of papers\cite{Nakanishi:1977gt,Nakanishi:1978ec,Nakanishi:1978zx,Nakanishi:1978np,Nakanishi:1979ff,Nakanishi:1980rf,Nakanishi:1980db}
 based on the Einstein-Hilbert action with BRS gauge fixing in 
de Donder gauge. It is summarized in his textbook\cite{Nakanishi:1990qm} 
co-authored with Ojima. 
He remarked there that the Einstein gravity field equation can also be 
rewritten 
in the form of Maxwell-type.\cite{Nakanishi:1981fj} 
In this formulation, he also found a 
beautiful theorem\cite{Nakanishi:1990qm} together with Ojima
that the graviton can be identified 
with a {\it Nambu-Goldstone (NG) massless tensor particle} accompanying 
the spontaneous breaking of $\GL(4)$ symmetry down to $SO(1,3)$ Lorentz 
symmetry, thus proving the exact masslessness of graviton in Einstein gravity 
theory.\footnote{Ogievetsky, independently, identified the graviton 
with the Nambu-Goldstone tensor in his non-linear realization theory for 
$\GL(4)/SO(1,3)$.\cite{Ogievetsky}} 
This is an 
gravitational extension of the Ferrari-Picasso's theorem\cite{Ferrari:1971at}
 which proves that 
the photon is a {\it NG vector boson} accompanying the spontaneous 
breaking of a vector-charge $Q_\mu$ symmetry, corresponding to the gauge 
symmetry with transformation parameter linear in $x^\mu$. 
Nakanishi also found in his $\GL(4)$-invariant de Donder gauge that there 
exists an $8+8$ dimensional Poincar\'e-like $\IOSp(8|8)$ supersymmetry which 
he called {\it choral symmetry} 
and contains (as its member) BRS and FP ghost scale symmetries as well as 
the $\GL(4)$ and rigid translation corresponding to the GC transformation with 
transformation parameter $\varepsilon^\mu$ linear in $x^\mu$, $\varepsilon^\mu=a^\mu_\nu x^\nu+ b^\mu$. 

However, this work is a formal theory based on the Einstein-Hilbert action. 
It is perturbative non-renormalizable and may not give a well-defined theory, 
although there is a possibility that it may satisfy the so-called asymptotic 
safety\cite{Niedermaier:2006wt}
 and gives a UV complete theory.\footnote{There recently appeared 
an interesting paper\cite{Pottel:2020iuz}
which proposes a novel perturbative approach to the 
Einstein-Hilbert gravity using the quadratic gravity terms as 
regulators which, the authors claim, can be removed eventually without harm.}
 
On the other hand, however, there are many investigations of higher derivative 
gravity theories. In particular, quadratic gravity\cite{Alvarez-Gaume:2015rwa,
Salvio:2014soa,Salvio:2018crh} 
attracted much attention 
in connection with the perturbative renormalizability\cite{Stelle:1976gc},
Weyl invariant theory\cite{Kaku:1977pa,Fradkin:1985am,%
Salvio:2017qkx},
and asymptotic freedom\cite{Julve:1978xn,Fradkin:1981iu,Holdom:2015kbf}.

These higher derivative theories suffer from 
the massive (negative metric) ghost problem in perturbative regime, 
although 
\erase{there have been many proposals for possible way out. 
This ghost problem is, however, out of scope of this paper.}
there have been many proposals for possible 
\correct{way out (See te.g.,}{ways out (See e.g.,} 
Ref.\cite{Salvio:2018crh} 
\add{for the review)}. 
This ghost problem is, however, out of \add{the} scope of this paper. 

\erase{
The quadratic gravity Lagrangian contains second order derivative metric 
field. 
From the effective field theory viewpoint, in which one regards 
the Einstein-Hilbert action as a low energy effective theory 
in the lowest derivative order, it is natural to consider even the action 
containing arbitrarily high order derivative fields.}

\add{Even if we are much less ambitious than to make gravity theory 
UV-complete, we still have several motivations to consider higher derivative 
 gravity theories.} 

\add{From the low energy effective field theory viewpoint, 
it is quite natural to consider the actions 
containing higher and higher order derivative fields, successively, 
from low to high energies. The 
Einstein-Hilbert action is the lowest derivative order, and 
the quadratic gravity actions are the next derivative order, 
and so on.} 

\add{Or, alternatively, one may simply wants a {\it gravity theory with 
a UV cut-off} $M$ valid only in the low energy region $E<M$. 
A simple momentum cut-off does not work here since it breaks the GC-invariance.
Pauli-Villars regulators respecting the GC-invariance can be supplied 
by considering the covariant higher derivative terms. As noted by 
Stelle, the gravity field propagator behaves like $\sim1/p^4$ in the quadratic 
gravity, and sufficiently cut-off the UV contribution to make the theory 
renormalizable in 4D.  
As regulators, however, to work sufficiently enough to make all the quantities 
finite in 4D, the propagator must drop as fast as $\sim1/p^6$. Such behavior 
would be supplied, for instance, by the quadratic 
term of covariant quantities which contain {\it third order 
derivatives of gravity field.} 
}

In this paper we will consider a general gravity theory which is 
invariant under the general coordinate (GC) transformation and contains 
arbitrarily high order derivatives of gravity and matter fields, and we 
\begin{enumerate}
\item derive a concrete form of the Noether current for the rigid translation, 
i.e., energy momentum tensor, and
\item derive the Maxwell-type gravity equation of motion in gauge unfixed, 
i.e., classical system, and 
\item the Maxwell-type equation analogous to Eq.~(\ref{eq:Maxwell-typeYMeom})
in gauge fixed quantum system in de Donder-Nakanishi gauge.
\item We also derive the Noether currents of the $\IOSp(8|8)$ symmetry, 
present in the de Donder-Nakanishi gauge.
\end{enumerate}

Original motivation for the present author to consider this problem is 
to give a sound proof for the existence theorem\cite{Kugo:1985jc}
 of massless graviton claiming 
that there should exist a spin 2 massless graviton in any GC invariant theory 
as far as it realizes a translational invariant vacuum with flat Minkowski 
metric. This is a generalization of the Ferrari-Picasso theorem for the massless photon and the Nakanishi-Ojima theorem for the massless graviton.
Those theorems were proved explicitly assuming the renormalizable QED and 
Einstein gravity, respectively. 
To prove the existence theorem generally, however, it is 
necessary to have Maxwell-type gravity equation of 
motion in any GC invariant system assuming no particular form of action.

This paper is organized as follows. In Section 2, we present 
totally general classical system containing arbitrarily high order 
derivative fields which is only assumed GC transformation invariant. 
To treat such a system, we introduce a series of generalized both-side 
derivatives and prove some 
formulas they satisfy. Based on them, we derive an expression for 
the energy-momentum tensor for such a general system as the Noether 
current for the translation invariance, and show that 
the gravity field equation of motion can be cast into the form of 
Maxwell-type equation. In Section 3, 
these results are generalized in the 
gauge-fixed system by adopting the $\GL(4)$-invariant de Donder gauge 
\`a la Nakanishi. In Section 4, using the same technique we show that 
each of Noether currents of the $\IOSp(8|8)$ symmetry can be written 
in a form of the source current of a Maxwell-type equation.
Section 5 is devoted to the conclusion. 
Some technical points on $\OSp$ transformations are discussed in 
Appendices A and B. In Appendix A, $\OSp(8|8)$ transformation of 
the gauge-fixing plus Faddeev-Popov term is computed for the 
$x^\mu$-dependent transformation parameter. 
In Appendix B, to get some familiarity with the $\OSp$-symmetry, 
we briefly study the simplest model, $\OSp(2|2)$-invariant scalar field 
system on flat Minkowski background; 
$\OSp(2|2)$ Noether current is derived and the $\IOSp(2|2)$ algebra 
are confirmed from the canonical (anti-)commutation relations.

\section{
Gravity Equation of Motion in a generic Higher Derivative 
System}

We consider a generic system whose action contains 
higher order derivative fields up to $N$-th order $\partial_{\soeji\mu_1N}$ 
\erase{($N$ may even be $\infty$)}:
$$ 
S[\phi] = \int{d^4}x \, {\cal L}\big(\phi,\ \partial_\mu\phi,\ 
  \partial_{\mu\nu}\phi, \ \partial_{\mu\nu\rho}\phi, \ \cdots, \partial_{\soeji\mu_1N}\phi 
\big)\ ,
$$
where $\phi$ stands for a collection of fields $\{\,\phi^j\,\}$ 
(whose index $i$ may be suppressed when unimportant) and 
we use abbreviations like
\begin{align}
\partial_{\mu_1\mu_2\cdots\mu_n} \phi^j
&\equiv\partial_{\mu_1}\partial_{\mu_2}\cdots\partial_{\mu_n}\phi^j\ , \\ 
{\cal L}_j^{\,;\mu_1\mu_2\cdots\mu_n} 
&\equiv{\partial{\cal L}\over\partial(\partial_{\mu_1\mu 
_2\cdots\mu_n}\phi^j) }
\Big|_{\rm weight\,1}  \ .
\end{align}%
The suffix `weight 1' in the latter means that we keep always the 
weight to be one irrespectively of whether the $n$ indices 
$\mu_1, \mu_2, \cdots, \mu_n$ take the same values or not; 
namely, for the case 
${\cal L}= a^{\mu\nu}\partial_{\mu\nu}\phi$, for instance, \  
$\partial{\cal L}/\partial(\partial_{11}\phi)= a^{11}$ and 
$\partial{\cal L}/\partial(\partial_{12}\phi)= a^{12}+a^{21}$,  but we 
\correct{{\em define}}{{\it define}}  
$\partial{\cal L}/\partial(\partial_{\mu\nu}\phi)
\big|_{\rm weight\, 1}= (a^{\mu\nu}+a^{\nu\mu})/2!$ 
\ always. 
The functional derivative of the action $S$ with respect to $\phi^j$ is
given by 
\begin{align}
{\delta S\over\delta\phi^j}
&={\partial{\cal L}\over\partial\phi^j}
-\partial_\mu{\cal L}_j^{\,;\mu}+\partial_{\mu\nu}{\cal L}_j^{\,; \mu\nu}-\partial_{\mu\nu\rho}{\cal L}_j^{\,;\mu\nu\rho}
+\cdots\ \nn
&=
\sum_{n=0}^N (-)^n \partial_{\soeji\mu_1n}{\cal L}_j^{\,;\soeji\mu_1n}, 
\label{eq:DerS}
\end{align}
where $N$ is the highest order $n$ of the derivative fields 
$\partial_{\soeji\mu_1n}\phi$ contained in ${\cal L}$ 
\add{(so that ${\cal L}_j^{\,;\soeji\mu_1n}=0$ for $n \geq N+1$)},  
and, for the $n=0$ case of empty set $\{\soeji\mu_1n\}$, 
$\partial_{\soeji\mu_1n}=1$ and ${\cal L}_j^{\,;\soeji\mu_1n}=\partial{\cal L}/\partial\phi^j$ are understood. 
The Euler-Lagrange equations are given by 
$\delta S/\delta\phi^j=0$.

The Lagrangian generally changes under 
an infinitesimal transformation 
$\phi\rightarrow\phi+ \delta\phi$, as 
\begin{align}
\delta{\cal L}&= 
{\partial{\cal L}\over\partial\phi^j}\delta\phi^j
+{\cal L}_j^{\,;\mu}\partial_\mu\delta\phi^j
+{\cal L}_j^{\,;\mu\nu}\partial_{\mu\nu}\delta\phi^j
+ \cdots
= \sum_{n=0}^N {\cal L}_j^{\,;\soeji\mu_1n}\partial_{\soeji\mu_1n}\delta\phi^j,
\label{eq:DerL}
\end{align}
where summation over the repeated $j$ is also implied. 
We consider the system which is invariant under the gauge transformation
taking the form
\begin{equation}
\delta\phi^j(x)= G^j_\rho\varepsilon^\rho(x)+ T^{j\mu}_{\ \ \rho} \partial_\mu\varepsilon^\rho(x).
\label{eq:gaugetrf}
\end{equation}
For the GC transformation $x^\rho\rightarrow x'^\rho=x^\rho-\varepsilon^\rho(x)$, 
this field transformation reads more explicitly\footnote{
Here, we are taking $\varepsilon^\rho$ with opposite sign to Nakanishi's so that 
the definitions of $G_\rho^j, T^{j\mu}_{\ \ \rho}$ and $\left[\phi^j\right]^\mu_\rho$ 
all have opposite sign to Nakanishi's.}
\begin{align}
&\delta\phi^j(x)= \varepsilon^\rho(x)\partial_\rho\phi^j+ \left[\phi^j\right]^\mu_\rho\partial_\mu\varepsilon^\rho(x) \nn
&\text{i.e.,}\ \  G^j_\rho=\partial_\rho\phi^j, \quad 
T^{j\mu}_{\ \ \,\rho}=\left[\phi^j\right]^\mu_\rho\,.
\label{eq:GCtrf}
\end{align}
For a general tensor field $\phi^j=T_{\nu_1\cdots\nu_p}{}^{\sigma_i\cdots\sigma_q}$,  
the symbol $\left[\phi^j\right]^\mu_\rho$ is defined by\cite{Nakanishi:1990qm}
\begin{equation}
\left[T_{\nu_1\cdots\nu_p}{}^{\sigma_1\cdots\sigma_q}\right]^\mu_\rho= 
\sum_{i=1}^p\delta^\mu_{\nu_i}T_{\nu_1\cdots\nu_{i-1}\rho\nu_{i+1}\cdots\nu_p}{}^{\sigma_1\cdots\sigma_q}
-\sum_{j=1}^q\delta_\rho^{\sigma_j}T_{\nu_1\cdots\nu_p}{}^{\sigma_1\cdots\sigma_{i-1}\mu\sigma_{i+1}\cdots\sigma_q}\,.
\end{equation}
The GC invariance of the system implies that the Lagrangian is a scalar 
density so that the change of ${\cal L}$ is given by a total divergence
\begin{equation}
\delta{\cal L}=\partial_\mu({\cal L}\varepsilon^\mu) \,. 
\label{eq:ChangeL}
\end{equation}
For 
the GC transformation $\delta\phi$ in Eq.~(\ref{eq:gaugetrf}),
we can equate this expression (\ref{eq:ChangeL}) for 
$\delta{\cal L}$ with Eq.~(\ref{eq:DerL}) and obtain an identity
\begin{align}
\sum_{n=0}^N \left[{\cal L}_j^{\,;\soeji\mu_1n}\partial_{\soeji\mu_1n}
\bigl(G^j_\rho\varepsilon^\rho(x)+ T^{j\mu}_{\ \ \rho} \partial_\mu\varepsilon^\rho(x)\bigr)\right]
-\partial_\mu({\cal L}\varepsilon^\mu)=0 \,.
\label{eq:DerLIdentity}
\end{align}
This equation, if expanded in a power of derivatives $\partial_\mu$ on the gauge 
transformation parameter (function) $\varepsilon^\rho(x)$, yields 
\begin{align}
&\left(\sum_{n=0}^N {\cal L}_j^{\,;\soeji\mu_1n}\partial_{\soeji\mu_1n}G^j_\rho 
-\partial_\rho{\cal L}\right) \varepsilon^\rho\nn
&+\left( \Bigl\{\sum_{n=0}^{N-1}(n+1){\cal L}_j^{\,;\mu\soeji\alpha_1n}
\partial_{\soeji\alpha_1n}G^j_\rho 
+\sum_{n=0}^N{\cal L}_j^{\,;\soeji\alpha_1n}\partial_{\soeji\alpha_1n}T^{j\mu}_{\ \ \rho}\Bigr\} 
-{\cal L}\delta^\mu_\rho\right) \partial_\mu\varepsilon^\rho\nn
&+\sum_{k=1}^{N} 
{\calK_k}^{\soeji\nu_1k \mu}{}_{\!\!\rho}\, \partial_{\soeji\nu_1k \mu}\varepsilon^\rho(x) =0 \ ,
\label{eq:expDerE}
\end{align}
where $\calK_k$'s $(k=0,1,2,\cdots)$ are defined by 
\begin{align}
{\calK_k}^{\soeji\nu_1k \mu}{}_{\!\!\rho}
\equiv 
&\sum_{n=0}^{N-k-1}
{}_{n+k+1}C_{k+1}\, 
{\cal L}_j^{\,;\soeji\nu_1k \mu\soeji\alpha_1n}\,\partial_{\soeji\alpha_1n}G^j_\rho 
\nn 
&+\sum_{n=0}^{N-k}
{}_{n+k}C_{k}\, 
{\cal L}_j^{\,;\soeji\nu_1k\soeji\alpha_1n}\,\partial_{\soeji\alpha_1n}
T^{j\mu}_{\ \ \rho} \ ,
\label{eq:defTk}
\end{align}
with ${}_nC_k$ denoting the binomial coefficient $\tbinom{n}{k}=n!/k!(n-k)!$. 
Note that the first summation term 
$\big\{\sum_{n=0}^{N-1}(\cdots)+\sum_{n=0}^N(\cdots)\big\}$ in the coefficient of 
$\partial_\mu\varepsilon^\rho$ in the second line in Eq.~(\ref{eq:expDerE}) 
is just identical with the quantity 
${\calK_k}^{\soeji\nu_1k \mu}{}_{\!\!\rho}$ for the case $k=0$. 

Since the functions $\varepsilon^\rho, \ \partial_\mu\varepsilon^\rho, \cdots, \partial_{\soeji\mu_1k}\varepsilon^\rho$ 
are mutually independent, the coefficients should vanish separately, 
implying the following $N+2$ identities which we shall refer to as 
{\it $\delta{\cal L}$-identities} below: 
\begin{align}
&\sum_{n=0}^N {\cal L}_j^{\,;\soeji\mu_1n}\partial_{\soeji\mu_1n}G^j_\rho-\partial_\rho{\cal L} =0 
\label{eq:L0}\\
&{\calK_0}^\mu_{\,\rho} -{\cal L}\delta^\mu_\rho=0
\label{eq:L1}\\
& 
\mathop{\text{Sym}}_{\{\soeji\nu_1k \mu\}}
\Bigl({\calK_k}^{\soeji\nu_1k \mu}{}_{\!\!\rho}\Bigr) =0 \qquad \text{for}\quad 
k=1,2,\cdots, N.
\label{eq:L2}
\end{align}
Here ${\mathop{\text{Sym}}}_{\{\soeji\nu_1k \mu\}}$ implies the totally symmetric part 
with respect to the $k+1$ indices $\nu_1,\cdots,\nu_k$ and $\mu$. Note that 
only the totally symmetric part of ${\calK_k}^{\soeji\nu_1k \mu}{}_{\!\!\rho}$ 
should vanish since it vanishes when multiplied by the totally symmetric 
function $\partial_{\soeji\nu_1k \mu}\varepsilon^\rho(x)$. Note also that 
${\calK_k}^{\soeji\nu_1k \mu}{}_{\!\!\rho}$ is manifestly symmetric with 
respect to the first $k$ indices $\nu_1,\cdots,\nu_k$ as is clear from the 
defining Eq.~(\ref{eq:defTk}).

Now we can derive useful identities for rewriting the suitable 
linear combination of equations of motion (\ref{eq:DerS}),\footnote{%
We call the quantities $\delta S/\delta\phi^j$ `equations of motion' although being 
an abuse of terminology since the equation of motion itself is 
the equation $\delta S/\delta\phi^j=0$.}
\begin{align}
{\delta S\over\delta\phi^j}
&=
\sum_{n=0}^N (-)^n \partial_{\soeji\mu_1n}{\cal L}_j^{\,;\soeji\mu_1n}\,.
\label{eq:DerSj}
\end{align}
First, a linear combination $-(\delta S/\delta\phi^j)G_\rho^j$ of the equations of motion 
is rewritten into the following form by adding the first 
$\delta{\cal L}$-identity (\ref{eq:L0}):
\begin{equation}
-(\delta S/\delta\phi^j)G_\rho^j=
\sum_{n=0}^N 
\left({\cal L}_j^{\,;\soeji\mu_1n}\,\partial_{\soeji\mu_1n}G^j_\rho 
- (-)^n \partial_{\soeji\mu_1n}{\cal L}_j^{\,;\soeji\mu_1n}\cdot G^j_\rho\right) -\partial_\rho{\cal L}\,. 
\label{eq:A16}
\end{equation}
To rewrite this more concisely, we introduce a generalized `both-side' 
derivative defined for $n\geq0$ by\cite{Hamazaki:1994rf}
\begin{align}
F\,
\bothDer{}{\longgLRarrow}{\mu_1\soeji\mu_2n} \,G
\ \equiv\ & 
F\partial_{\mu_1\mu_2\cdots\mu_n}G
- \partial_{\mu_1}F\cdot \partial_{\mu_2\cdots\mu_n}G \nn
 & + \partial_{\mu_1\mu_2}F\cdot \partial_{\mu_3\cdots\mu_n}G
  - \cdots+(-)^n\partial_{\mu_1\mu_2\cdots\mu_n}F\cdot G \ , 
\label{eq:bothDer0}
\end{align}
for arbitrary two functions $F$ and $G$, with understanding 
$\bothDer{}{\longgLRarrow}{\mu_1\soeji\mu_2n}=1$ when $n=0$. 
This derivative is no longer symmetric under permutation of 
the indices but satisfies a useful formula\cite{Hamazaki:1994rf}
\begin{equation}
\partial_\mu\big[ F^{\mu\alpha_1\cdots\alpha_n}
\bothDer{}{\longLRarrow}{\soeji\alpha_1n}
\,G\big] 
= F^{\mu\alpha_1\cdots\alpha_n}\partial_{\mu\alpha_1\cdots 
\alpha_n}G 
+ (-)^n\partial_{\mu\alpha_1\cdots\alpha_n}F^{\mu\alpha_1\cdots 
\alpha_n}\cdot G \ , 
\label{eq:F0}%
\end{equation} 
for any totally symmetric function $F^{\mu\alpha_1\cdots\alpha_n}$ 
with respect to the $n+1$ indices 
$\{\, \mu, \alpha_1, \cdots, \alpha_n\,\}$.  Applying this formula we can 
rewrite the identity (\ref{eq:A16}) as 
\begin{align}
-{\delta S\over\delta\phi^j}G^j_\rho&= \partial_\mu J^\mu_{\ \ \rho} \nn
J^\mu_{\ \rho} &= \sum _{n=0}^{N-1}
{\cal L}_j^{\,;\mu\soeji\alpha_1n}
\bothDer{}{\longLRarrow}{\soeji\alpha_1n}G^j_\rho- \delta^\mu_\rho{\cal L}\ .
\label{eq:NoetherCurrent}
\end{align}
This $J^\mu_{\ \rho}$ is the Noether current for the global GC transformation 
with $x$-independent $\varepsilon^\rho$ ($=$ translation), i.e., energy-momentum tensor, 
for the higher derivative 
system. This identity shows that it is indeed conserved when the equations 
of motion $\delta S/\delta\phi^j=0$ are satisfied. 

Now in order to derive various identities from the rest of the 
$\delta{\cal L}$-identities, (\ref{eq:L1}) and (\ref{eq:L2}), we need to introduce 
generalized both-side derivatives and some formulas for them.  

We define $k$-th both-side derivative $\stackrel{k}{\longleftrightarrow}$ 
by induction both in the number $k$ and the differential order $n$: 
\begin{align}
&\bothDer{k}{\longLRarrow}{\soeji\alpha_1n} \equiv 
\bothDer{k}{\longLRarrow}{\soeji\alpha_1{n-1}}\cdot\partial_{\alpha_n} + 
\bothDer{k-1}{\longLRarrow}{\soeji\alpha_1n} 
\label{eq:Def_k-bothDer}
\\ 
&\text{with initial condition} \nn
& 
\left\{
\begin{array}{rcl}
k=-1    &:& \quad \bothDer{-1}{\longLRarrow}{\soeji\alpha_1n} 
= (-)^n \bothDer{}{\longLarrow}{\soeji\alpha_1n}\ \  \text{for}\ \ \forall n\geq0 \\        
n=0    &:& \quad \bothDer{k}{\longLRarrow}{\soeji\alpha_1n}\Big|_{n=0} = 1 
\ \  \text{for}\ \ \forall k\geq-1         
\end{array}
\right. \ .
\label{eq:InitCond}
\end{align}
It is easy to see that the 
$\stackrel{k=0}{\longleftrightarrow}$ is just the same as the original 
`both-side' derivative $\longleftrightarrow$ introduced above 
in Eq.~(\ref{eq:bothDer0}); 
indeed, it satisfies the above recursive defining relation (\ref{eq:Def_k-bothDer}) for $k=0$ as follows:
\begin{align}
\bothDer{}{\longLRarrow}{\soeji\alpha_1n}
&=
\sum_{\ell=0}^{n} (-)^\ell
\bothDer{}{\longLarrow}{\soeji\alpha_1\ell}\cdot
\bothDer{}{}{\soeji\alpha_{\ell+1}n} \nn
&=\sum_{\ell=0}^{n-1} 
(-)^\ell\Bigl(\bothDer{}{\longLarrow}{\soeji\alpha_1{\ell}}\cdot
\bothDer{}{}{\soeji\alpha_{\ell+1}{n-1}}\Bigr)\partial_{\alpha_n}
+(-)^n \bothDer{}{\longLarrow}{\soeji\alpha_1n} \nn
&=
\bothDer{}{\longLRarrow}{\soeji\alpha_1{n-1}}\cdot\partial_{\alpha_n} + 
\bothDer{k=-1}{\longLRarrow}{\soeji\alpha_1n} \ .
\end{align}
Then, as a generalization of the $k=0$ formula (\ref{eq:F0}), 
we have the following 
formula which holds for all $k\geq0, n\geq0$ and for any 
totally symmetric function $F^{\mu\alpha_1\cdots\alpha_n}$ with respect to the 
$n+1$ indices $\{\,\mu, \alpha_1,\cdots,\alpha_n\,\}$:
\begin{equation}
\partial_{\mu}\Big[ F^{\mu\alpha_1\cdots\alpha_n}
\bothDer{k}{\longLRarrow}{\soeji\alpha_1n}
\,G\Big] 
= F^{\mu\alpha_1\cdots\alpha_n}\Bigl( {}_{n+k+1}C_k \,\partial_{\mu\alpha_1\cdots\alpha_n} 
- \bothDer{k-1}{\longLRarrow}{\mu\soeji\alpha_1n}\Bigr) \,G\ .
\label{eq:Fk}%
\end{equation}
The proof easily goes by induction in the number $N\equiv k+n$ in the region 
$k\geq0$ and $n\geq0$. First note that this formula holds 
at $k=0$ boundary as shown above for $\forall n\geq0$, and clearly 
hold also at $n=0$ boundary with $\forall k\geq0$ since the relevant $k$-th both-side 
derivatives there are just 
$\bothDer{k}{\longLRarrow}{\soeji\alpha_1n}\bigr|_{n=0}=1$
and that of a single derivative 
$\overset{\stackrel{k-1}{\longleftrightarrow}}{\partial_\mu}$ 
which is simply, by Eq.~(\ref{eq:Def_k-bothDer}),
\begin{equation}
\overset{\stackrel{k-1}{\longleftrightarrow}}{\partial_\mu}
=\partial_\mu+\overset{\stackrel{k-2}{\longleftrightarrow}}{\partial_\mu}
=k\partial_\mu+\overset{\stackrel{-1}{\longleftrightarrow}}{\partial_\mu}
=k\partial_\mu-\overset{\leftarrow}{\partial_\mu}\,.
\end{equation}
So it is sufficient to prove the formula only for $k\geq1$ and $n\geq1$. 
Now assume that the formula (\ref{eq:Fk}) holds for all 
$k\geq0$ and $n\geq0$ values 
in the region $k+n\leq\exists N$, and let us evaluate the LHS of the formula for 
any $k\geq1$ and $n\geq1$ with $k+n=N+1$. If we use the defining 
Eq.~(\ref{eq:Def_k-bothDer}) 
\begin{align}
F^{\mu\alpha_1\cdots\alpha_n}\bothDer{k}{\longLRarrow}{\soeji\alpha_1n}\,G 
&=F^{\mu\alpha_1\cdots\alpha_n}\bothDer{k}{\longLRarrow}{\soeji\alpha_1{n-1}}(\partial_{\alpha_n}G) 
+ F^{\mu\alpha_1\cdots\alpha_n}\bothDer{k-1}{\longLRarrow}{\soeji\alpha_1n}\,G \ ,
\end{align}
the two terms on the RHS have lower values $k+n=N$ by one than the LHS, 
to which we can apply the formula by the induction assumption, so that
\begin{align}
\partial_{\mu}\Big[ F^{\mu\alpha_1\cdots\alpha_n}\bothDer{k}{\longLRarrow}{\soeji\alpha_1n}\,G\Big] 
&= F^{\mu\alpha_1\cdots\alpha_n}\Bigl( {}_{n+k}C_k \,\partial_{\mu\alpha_1\cdots\alpha_{n-1}} 
- \bothDer{k-1}{\longLRarrow}{\mu\soeji\alpha_1{n-1}}\Bigr) (\partial_{\alpha_n}G) \nn
&+ F^{\mu\alpha_1\cdots\alpha_n}\Bigl( {}_{n+k}C_{k-1} \,\partial_{\mu\alpha_1\cdots\alpha_n} 
- \bothDer{k-2}{\longLRarrow}{\mu\soeji\alpha_1n}\Bigr) \,G \nn
&= F^{\mu\alpha_1\cdots\alpha_n}\Bigl[\bigl( {}_{n+k}C_k +{}_{n+k}C_{k-1} \bigr)
\,\partial_{\mu\alpha_1\cdots\alpha_n}\,G  \nn
&- \Bigl(
\bothDer{k-1}{\longLRarrow}{\mu\soeji\alpha_1{n-1}}\,\partial_{\alpha_n} +
 \bothDer{k-2}{\longLRarrow}{\mu\soeji\alpha_1n}\Bigr)\,G \Bigr]  \ .
\end{align}
If we note an identity (of Pascal's triangle) 
${}_{n+k}C_k+{}_{n+k}C_{k-1}={}_{n+k+1}C_k$ and apply 
again the defining Eq.~(\ref{eq:Def_k-bothDer}) with $k\rightarrow k-1$, then 
we see that the last expression is just reproducing the RHS of 
the formula (\ref{eq:Fk}), finishing the proof. 

Now we are ready to derive the Maxwell-type form of gravity equation of motion. 
For that purpose let us introduce the following quantity $\calJ_k$ 
for $k=0, 1,2, \cdots$: 
\begin{align}
{\calJ_k}^{\soeji\nu_1k \mu}{}_{\!\!\rho}
\equiv 
&\sum_{n=0}^{N-k-1}
{\cal L}_j^{\,;\soeji\nu_1k\mu\soeji\alpha_1n}\,
\bothDer{k}{\longLRarrow}{\soeji\alpha_1n}\,G^j_\rho 
\nn 
&
+\sum_{n=0}^{N-k}{\cal L}_j^{\,;\soeji\nu_1k\soeji\alpha_1n}\,
\bothDer{k-1}{\longLRarrow}{\soeji\alpha_1n}\,T^{j\mu}_{\ \,\rho} \ .
\label{eq:defJk}
\end{align}
The first of this quantity ${\calJ_0}^\mu_{\,\rho}$ with $k=0$ is a combination of 
equation of motion, $(\delta S/\delta\phi^j)T^{j\mu}_{\ \,\rho}$, Lagrangian ${\cal L}$ 
and the energy-momentum tensor $J^\mu_{\,\rho}$:
\begin{equation}
{\calJ_0}^\mu_{\ \rho}= 
J^\mu_{\,\rho}+\delta^\mu_\rho{\cal L} + 
\frac{\delta S}{\delta\phi^j}T^{j\mu}_{\ \,\rho}\,.
\label{eq:calJ0}
\end{equation}
This can be seen from Eqs.~(\ref{eq:NoetherCurrent}) and (\ref{eq:DerSj}) which 
 are rewritten by using the definition 
 of the $k$-th both-side derivative with $k=0$ and $-1$, respectively, into 
\begin{align}
\sum_{n=0}^{N-1}{\cal L}_j^{\,;\mu\soeji\alpha_1n}\,
\bothDer{0}{\longLRarrow}{\soeji\alpha_1n}\,G^j_\rho 
&=J^\mu_{\,\rho}+\delta^\mu_\rho{\cal L}  
\nn 
\sum_{n=0}^{N}{\cal L}_j^{\,;\soeji\alpha_1n}\,
\bothDer{-1}{\longLRarrow}{\soeji\alpha_1n}\,T^{j\mu}_{\ \,\rho} 
&=\frac{\delta S}{\delta\phi^j}T^{j\mu}_{\ \,\rho}\,.
\end{align}

Owing to the general formula (\ref{eq:Fk}), the two quantities, 
$\calJ_k$ introduced here (\ref{eq:defJk}) and 
$\calK_k$ defined previously in Eq.~(\ref{eq:defTk}), 
satisfy the following recurrence relation:
\begin{equation}
{\calK_k}^{\soeji\nu_1k\mu}{}_{\!\!\rho}
-{\calJ_k}^{\soeji\nu_1k\mu}{}_{\!\!\rho}
=\partial_\tau{\calJ_{k+1}}^{\tau\soeji\nu_1k\mu}{}_{\!\!\rho}\,.
\label{eq:RecRel}
\end{equation}
When applying the formula (\ref{eq:Fk}) to derive this equality,  
we should note that the summation 
over the set of $n+1$ dummy indices $\{ \tau, \soeji\alpha_1n \}$ contained in 
the RHS   quantity 
$\partial_\tau{\calJ_{k+1}}^{\tau\soeji\nu_1k\mu}{}_{\!\!\rho}$ is identified with 
the summation over the set $\{ \soeji\alpha_1{n+1} \}$ contained 
in the LHS quantities 
${\calK_k}^{\soeji\nu_1k\mu}{}_{\!\!\rho}
-{\calJ_k}^{\soeji\nu_1k\mu}{}_{\!\!\rho}$
by identifying $\alpha_{n+1}$ as $\tau$. This implies that the $n=0$ terms 
existing in the summations $\sum_{n=0}$ in $\calK_k$ and $\calJ_k$ on the LHS 
do not appear on the RHS. However, the $n=0$ terms in 
$\calK_k$ and $\calJ_k$ are seen to be the same, so 
canceling themselves on the LHS.  

From this relation (\ref{eq:RecRel}), we find, suppressing the tensor indices, 
\begin{align}
{\calJ_k}&={\calK_k}-\partial{\calJ_{k+1}}
={\calK_k}-\partial{\calK_{k+1}}+\partial^2{\calJ_{k+2}} \nn
&= \cdots 
=\sum_{\ell=0}^K(-\partial)^\ell{\calK_{k+\ell}}+(-\partial)^{K+1}{\calJ_{k+K+1}}\ .
\end{align}
Since ${\cal L}^{\,;\soeji\nu_1k}=0$ for $k>N$, $\calJ_k$ as well as $\calK_k$ 
vanish for $k\geq N+1$. So we find the following expression for 
${\calJ_0}^\mu_{\,\rho}$ reviving the tensor indices:
\begin{equation}
{\calJ_0}^\mu{}_{\!\!\rho}
=\sum_{k=0}^N (-)^k\partial_{\soeji\nu_1k}{\calK_k}^{\soeji\nu_1k\mu}{}_{\!\!\rho}\ .
\label{eq:J0=T}
\end{equation}
We now insert Eq.~(\ref{eq:calJ0}) into $\calJ_0$ on the LHS, then, 
noting that 
the term $\delta^\mu_\rho{\cal L}$ there cancels the ${\calK_{k=0}}$ term on the RHS 
due to the second $\delta{\cal L}$-identity (\ref{eq:L1}), 
${\calK_0}^\mu_{\,\rho}=\delta^\mu_\rho{\cal L}$,
we find the gravity field equation in the form 
\begin{align}
\frac{\delta S}{\delta\phi^j}T^{j\mu}_{\ \,\rho}
&= -J^\mu_{\,\rho} 
+\sum_{k=1}^N (-)^k\partial_{\soeji\nu_1k}{\calK_k}^{\soeji\nu_1k\mu}{}_{\!\!\rho}\ .
\label{eq:GravityEq0}
\end{align} 
This is still not the final form. The last summation term can be written as 
a divergence form of a `field-strength' tensor 
$\calF^{\nu\mu}_{\ \ \rho}$, but it is not yet 
$\nu\mu$ antisymmetric:
\begin{align}
&\sum_{k=1}^N (-)^k\partial_{\soeji\nu_1k}{\calK_k}^{\soeji\nu_1k\mu}{}_{\!\!\rho}
\equiv- \partial_\nu\calF^{\nu\mu}_{\ \ \rho} 
\nn 
&\calF^{\nu\mu}_{\ \ \rho}
=
\sum_{k=0}^{N-1} (-)^k\partial_{\soeji\nu_1k}{\calK_{k+1}}^{\soeji\nu_1k\nu\mu}{}_{\!\!\rho}
 \ .
\label{eq:Fmunu}
\end{align}
However, thanks to the remaining $\delta{\cal L}$-identities (\ref{eq:L2}), we can 
modify it into an $\nu\mu$ antisymmetric field strength 
$\widetilde\calF^{\nu\mu}_{\ \ \rho}$ 
satisfying 
\begin{equation}
\partial_\nu\calF^{\nu\mu}_{\ \ \rho} =
\partial_\nu\widetilde\calF^{\nu\mu}_{\ \ \rho} \,.
\end{equation}

As noted before, the tensor ${\calK_k}^{\soeji\nu_1k\mu}{}_{\!\!\rho}$ 
defined in Eq.~(\ref{eq:defTk}) is manifestly totally symmetric with respect to 
the first $k$-indices $\{\nu_1, \cdots, \nu_k\}$. 
The $\delta{\cal L}$-identities (\ref{eq:L2}) say that it vanishes if further 
symmetrized including the last index $\mu$; namely, taking the cyclic 
permutation of the $k+1$ indices $\{\nu_1, \cdots, \nu_k, \mu\}$
\begin{equation}
{\calK_k}^{\soeji\nu_1k\mu}{}_{\!\!\rho}
+{\calK_k}^{\soeji\nu_2k \mu\nu_1}{}_{\!\!\rho}
+{\calK_k}^{\soeji\nu_3k \mu\nu_1\nu_2}{}_{\!\!\rho}
+ \cdots 
+{\calK_k}^{\mu\soeji\nu_1k}{}_{\!\!\rho}
=0 \,.
\end{equation}
If we act $k$-ple divergence $\partial_{\soeji\nu_1k}$ on this, 
the $\nu_j$ indices become dummy, and, since the manifest total symmetry among 
the first $k$ indices of $\calK_k$,
the $k$ terms from the second to the last yield the same quantity and we get 
\begin{equation}
\partial_{\soeji\nu_1k}{\calK_k}^{\soeji\nu_1k\mu}{}_{\!\!\rho}
+k\,\partial_{\soeji\nu_1k}{\calK_k}^{\soeji\nu_2k \mu\nu_1}{}_{\!\!\rho}
=0 \, .
\end{equation}
Or, taking $k\rightarrow k+1$ and renaming $\nu_{k+1}\rightarrow\nu$ in the first term and 
$\nu_1\rightarrow\nu$ 
in the second term, we have
\begin{equation}
\partial_\nu\left(\partial_{\soeji\nu_1k}\calK_{k+1}^{\ \, \soeji\nu_1k\nu\mu}{}_{\!\!\rho}
+(k+1)\,\partial_{\soeji\nu_1k}\calK_{k+1}^{\ \, \soeji\nu_1k \mu\nu}{}_{\!\!\rho}\right)
=0 \, ..
\label{eq:zerodiv}
\end{equation}
This means that $\calK_{k+1}^{\ \,\soeji\nu_1k\nu\mu}{}_{\!\!\rho}$ can be 
replaced by a $\nu\mu$ anti-symmetric tensor which we can define as
\begin{equation}
\widetilde\calK_{k+1}^{\ \,\soeji\nu_1k\nu\mu}{}_{\!\!\rho}
=
\frac{k+1}{k+2}\left(\calK_{k+1}^{\ \,\soeji\nu_1k\nu\mu}{}_{\!\!\rho}
-\calK_{k+1}^{\ \,\soeji\nu_1k\mu\nu}{}_{\!\!\rho}\right) \ .
\label{eq:defTildeTk}
\end{equation}
Indeed the difference between $\calK_{k+1}$ and $\widetilde\calK_{k+1}$ is 
given by 
\begin{equation}
\calK_{k+1}^{\ \,\soeji\nu_1k\nu\mu}{}_{\!\!\rho}
-\widetilde\calK_{k+1}^{\ \,\soeji\nu_1k\nu\mu}{}_{\!\!\rho}
=\frac1{k+2}
\left(\calK_{k+1}^{\ \,\soeji\nu_1k\nu\mu}{}_{\!\!\rho}
+(k+1)\,\calK_{k+1}^{\ \,\soeji\nu_1k \mu\nu}{}_{\!\!\rho}\right) \ ,
\end{equation}
whose $(k+1)$-ple divergence $\partial_\nu\partial_{\soeji\nu_1k}$ is guaranteed to 
vanish by Eq.~(\ref{eq:zerodiv}). 
Thus we find that the `field-strength' $\calF^{\nu\mu}_{\ \ \rho}$ in 
Eq.~(\ref{eq:Fmunu}) can be replaced by the $\nu\mu$ anti-symmetric one:
\begin{equation}
\widetilde\calF^{\nu\mu}_{\ \ \rho}
=
\sum_{k=0}^{N-1} (-)^k\partial_{\soeji\nu_1k}{\widetilde\calK_{k+1}}^{\soeji\nu_1k\nu\mu}{}_{\!\!\rho} \, .
\label{eq:fieldstrength}
\end{equation} 
With this antisymmetric field strength,
the gravity equation of motion is finally written in 
the desired form of the Maxwell-type equation:
\begin{align}
-\frac{\delta S}{\delta\phi^j}T^{j\mu}_{\ \,\rho}
&= J^\mu_{\,\rho} - \partial_\nu\widetilde\calF^{\mu\nu}_{\ \ \rho} \, .
\label{eq:Maxwell-typeGEq}
\end{align}
This is an equation for the gauge-unfixed classical system.   

Here, we note a more explicit expression for 
$\widetilde\calK_{k+1}$ in terms of the Lagrangian. 
Substituting the expression (\ref{eq:defTk}) for $\calK_k$ into 
the definition (\ref{eq:defTildeTk}), 
we note that the $G^j_\rho$-proportional part contained in 
${\calK_{k+1}}^{\soeji\nu_1k\nu\mu}$ is $\nu\mu$ symmetric so that only the 
$T^{j\mu}_{\ \ \rho}$-proportional part contributes to $\widetilde\calK_{k+1}$, and obtain
\begin{align}
{\widetilde\calK_{k+1}}^{\soeji\nu_1k \nu\mu}{}_{\!\!\rho}
&\equiv 
\frac{k+1}{k+2}
\sum_{n=0}^{N-k-1}
{}_{n+k+1}C_{k+1}\, 
\left(
{\cal L}_j^{\,;\soeji\nu_1k\nu\soeji\alpha_1n}\,\partial_{\soeji\alpha_1n}T^{j\mu}_{\ \ \rho} 
- (\nu\leftrightarrow \mu)
\right) \nn
&\text{for}\quad k=0, 1, 2, \cdots\ .
\label{eq:TildeTk}
\end{align}

\section{Quantum theory with de Donder gauge}

Let us now consider the quantum system. 
We add the gauge-fixing and corresponding Faddeev-Popov(FP) term to 
the classical GC invariant Lagrangian ${\cal L}_{\rm cl}$. (We call the 
Lagrangian ${\cal L}$ in the previous section ${\cal L}_{\rm cl}$ hereafter.)
We actually adopt Nakanishi's simpler form of 
${\cal L}_{\GF}+{\cal L}_{\FP}={\cal L}_{\GFP}$\cite{Nakanishi:1977gt,Nakanishi:1978ec} 
\begin{eqnarray}
{\cal L}_{\GFP}
&=& h \BrsN(i\kappa^{-1}g^{\mu\nu}\partial_\mu\bar c_\nu) 
= -\kappa^{-1}\tilde g^{\mu\nu}\partial_\mu b_\nu 
-i\tilde g^{\mu\nu}\cdot\partial_\mu\bar c_\rho\cdot\partial_\nu c^\rho 
\nn
&=&
\Brs(i\kappa^{-1}\tilde g^{\mu\nu}\partial_\mu\bar c_\nu) - 
 \partial_\mu\left( 
 i\tilde g^{\lambda\nu}\partial_\lambda\bar c_\nu\cdot c^\mu\right), 
 \label{eq:NakanishiGF}
\end{eqnarray}
with $h\equiv\sqrt{-g}$ and $\tilde g^{\mu\nu}\equiv hg^{\mu\nu}$. 
Here the usual BRS 
transformation $\Brs$ (obtained by replacing 
$-\varepsilon^\mu(x) \rightarrow \kappa c^\mu(x)$ for the usual gravity/matter fields) is given 
by a sum of 
Nakanishi's BRS $\BrsN$ and the translation $-\kappa c^\lambda\partial_\lambda$:
\begin{eqnarray}
\Brs\Phi&=& \BrsN\Phi- \BrsN(x^\lambda)\partial_\lambda\Phi, \nn
\BrsN(x^\lambda)&=& \kappa c^\lambda, \qquad \BrsN\Phi= -\kappa\partial_\mu c^\alpha\cdot \bigl[\Phi\bigr]^\mu_\alpha, \nn
\Brs{\tilde g^{\mu\nu}}&=& 
\kappa\left(\partial_\lambda c^\mu\cdot\tilde g^{\lambda\nu}+\partial_\lambda c^\nu\cdot\tilde g^{\mu\lambda} 
-\partial_\lambda(c^\lambda\tilde g^{\mu\nu})\right) \nn
\Brs\bar c_\mu&=& i B_\mu, \qquad 
\BrsN\bar c_\mu= i b_\mu, \qquad 
B_\mu= b_\mu+ i\kappa c^\lambda\partial_\lambda\bar c_\mu, \nn
\BrsN c^\mu&=&0,\qquad \Brs c^\mu= -\kappa c^\lambda\partial_\lambda c^\mu. 
\end{eqnarray}
We call this gauge specified by the gauge-fixing and FP term 
${\cal L}_\GFP$ in (\ref{eq:NakanishiGF}) ``de Donder-Nakanishi gauge". 
It corresponds to the 
de Donder-Landau gauge possessing no $\alpha\eta^{\mu\nu}B_\mu B_\nu$ term 
violating $\GL(4)$ invariance by the use of $\eta^{\mu\nu}$. 
Since the present ${\cal L}_\GFP$ for the de Donder-Nakanishi gauge is given in  a usual BRS exact form for the de Donder-Landau gauge up to a 
total derivative term as shown 
\add{in the last expression} in Eq.~(\ref{eq:NakanishiGF}), 
it is also invariant under the usual BRS transformation $\Brs$. 
The use of Nakanishi's BRS $\BrsN$, which represents the tensorial 
transformation part of the usual BRS transformation $\Brs$, and the use 
of the $b_\mu=i^{-1}\BrsN \bar c_\mu$ field, in particular, make manifest the 
existence of much larger $\IOSp(8|8)$ symmetry, called {\it choral symmetry} 
by Nakanishi, which contains symmetries of energy-momentum, $\GL(4)$, BRS, 
FP-ghost scale transformation etc as will be discussed explicitly 
in the next section.

We still consider the GC transformation $x^\rho\rightarrow x'^\rho=x^\rho-\varepsilon^\rho(x)$ 
in this quantum theory to derive identities. 
The gravity/matter fields $\phi^j$ 
are transformed in the 
same way as before:
\begin{align}
&\delta\phi^j(x)= G^j_\rho\varepsilon^\rho(x)+ T^{j\mu}_{\ \ \rho} \partial_\mu\varepsilon^\rho(x),
&\text{with}\ \  G^j_\rho=\partial_\rho\phi^j, \quad 
T^{j\mu}_{\ \ \,\rho}=\left[\phi^j\right]^\mu_\rho\ .
\label{eq:GCtrf0}
\end{align}
We call the newly added fields $b_\mu,\ \bar c_\mu$ and $c^\mu$ 
ghost fields collectively, and treat them all as scalar fields 
under GC transformation; namely denoting ghost fields by 
$\phi^M = (b_\mu, \bar c_\mu, c^\mu)$ collectively, 
\begin{align}
&\delta\phi^M(x)= G^M_\rho\varepsilon^\rho(x)+ T^{M\mu}_{\ \ \ \rho} \partial_\mu\varepsilon^\rho(x),
&\text{with}\ \  G^M_\rho=\partial_\rho\phi^M, \quad 
T^{M\mu}_{\ \ \ \rho}=0 \, .
\label{eq:GCtrf1}
\end{align}
Of course, ${\cal L}_\GFP$ is not invariant under the GC transformation, 
but we can easily calculate the change by noting the structure of 
the ${\cal L}_\GFP$, which is written formally as a scalar density:
\begin{equation}
{\cal L}_\GFP
= -\kappa^{-1}\tilde g^{\mu\nu}E_{\mu\nu}, \quad E_{\mu\nu}=\partial_\mu b_\nu 
+i\kappa\cdot\partial_\mu\bar c_\rho\cdot\partial_\nu c^\rho.
\end{equation}   
If the ghost part tensor $E_{\mu\nu}$ truly behaved as a $\mu\nu$ covariant tensor, 
${\cal L}_\GFP$ were a scalar density transforming only into the total divergence 
$\partial_\mu({\cal L}_\GFP \varepsilon^\mu)$. 
This is actually true for the FP ghost part 
$i\kappa\cdot\partial_\mu\bar c_\rho\cdot\partial_\nu c^\rho$ in $E_{\mu\nu}$ 
since $c_\rho$ and $\bar c^\rho$ are regarded as 
scalars so that their simple derivatives $\partial_\mu\bar c_\rho$ and 
$\partial_\nu c^\rho$ behave as $\mu$ and $\nu$ vectors, giving the desired $\mu\nu$ tensor 
as a product. 
But the NL field part $\partial_\mu b_\nu$ transforms just as a $\mu$ vector 
since $b_\nu$ is regarded as a scalar, so that the $\nu$ leg rotation part of  
the transformation of $\tilde g^{\mu\nu}$, i.e., 
$\delta\tilde g^{\mu\nu}\supset-\tilde g^{\mu\rho}\partial_\rho\varepsilon^\nu$, is not canceled. 
We thus see 
\begin{equation}
\delta{\cal L}_\GFP
= \kappa^{-1} \tilde g^{\mu\rho}\partial_\mu b_\nu\cdot \partial_\rho\varepsilon^\nu+ \partial_\mu({\cal L}_\GFP\varepsilon^\mu)\ . 
\label{eq:LGFPchange}
\end{equation}
So, the total Lagrangian in our quantum gravity theory
\begin{equation}
{\cal L}= {\cal L}_{\rm cl} + {\cal L}_\GFP
\end{equation}
changes under the GC transformation as 
\begin{equation}
\delta{\cal L}= \partial_\mu({\cal L}\varepsilon^\mu) + \kappa^{-1} \tilde g^{\mu\rho}\partial_\mu b_\nu\cdot \partial_\rho\varepsilon^\nu 
= \partial_\rho{\cal L}\cdot\varepsilon^\rho+ ({\cal L}\delta^\mu_\rho+\kappa^{-1} \tilde g^{\mu\nu}\partial_\nu b_\rho)\partial_\mu\varepsilon^\rho\ .
\end{equation}
Namely, this differs from Eq.~(\ref{eq:ChangeL}) in the classical system case 
only in the point that 
the $\kappa^{-1}\tilde g^{\mu\nu}\partial_\nu b_\rho$ term is added in the first order 
derivative term $\propto\partial_\mu\varepsilon^\rho$. 
Therefore, the $\delta{\cal L}$-identities in the previous section 
almost all remain the same and only the first order $\propto\partial_\mu\varepsilon^\rho$ 
identity (\ref{eq:L1}) is slightly changed into
\begin{equation}
{\calK_0}^\mu_{\,\rho} - ({\cal L}\delta^\mu_\rho+\kappa^{-1} \tilde g^{\mu\nu}\partial_\nu b_\rho) =0 \,.
\label{eq:L1mod}
\end{equation}
Note that we should now understand that 
that ${\cal L}$ is the total Lagrangian containing the ghost part 
${\cal L}_\GFP$ also and the fields $\phi^j$ cover not only the gravity/matter 
fields $\phi^j$ but also the ghost fields $\phi^M=(b_\mu,\bar c_\mu, c^\mu)$. 
The equation of motion $\delta S/\delta\phi$, of course, takes 
the same form (\ref{eq:DerSj}) as before. 
The zeroth order $\delta{\cal L}$-identity (\ref{eq:L0}), in particular, 
remains the same and the 
global translation current (Energy-momentum tensor) is given by the same 
form equation as Eq.~(\ref{eq:NoetherCurrent}):
\begin{align}
-{\delta S\over\delta\phi^j}G^j_\rho&= \partial_\mu J^\mu_{\ \ \rho} \nn
J^\mu_{\ \rho} &= \sum _{n=0}^{N-1}
{\cal L}_j^{\,;\mu\soeji\alpha_1n}
\bothDer{}{\longLRarrow}{\soeji\alpha_1n}G^j_\rho- \delta^\mu_\rho{\cal L}\ .
\label{eq:GCNoetherCurrentQuantum}
\end{align}
So Eq.~(\ref{eq:calJ0}) for $\calJ_0$ holds unchanged. The identity 
(\ref{eq:J0=T}) also holds as it stands. In going from Eq.~(\ref{eq:J0=T}) 
to the gravity equation (\ref{eq:GravityEq0}), however, 
the ${\cal L}\delta^\mu_\rho$ term from $\calJ_0$ now does not totally cancel the first 
term ${\calK_{k=0}}^\mu_{\,\rho}={\cal L}\delta^\mu_\rho+\kappa^{-1} \tilde g^{\mu\nu}\partial_\nu b_\rho$  
but leaves the $\kappa^{-1} \tilde g^{\mu\nu}\partial_\nu b_\rho$ term.  
Thus the Eq.~(\ref{eq:GravityEq0}) is now replaced by 
\begin{align}
\frac{\delta S}{\delta\phi^j}T^{j\mu}_{\ \,\rho}
&= -J^\mu_{\,\rho} 
+\kappa^{-1} \tilde g^{\mu\nu}\partial_\nu b_\rho 
+\sum_{k=1}^N (-)^k\partial_{\soeji\nu_1k}{\calK_k}^{\soeji\nu_1k\mu}{}_{\!\!\rho}\, .
\label{eq:QGravityEq0}
\end{align}
Note here that the implicit summation over $\phi^j$ also contains 
the ghost fields $\phi^M$ which contribute only to
the $n=0$ terms since the ghost fields appear only in the first order 
derivatives in the de Donder-Nakanishi gauge Lagrangian (\ref{eq:NakanishiGF}). 
 
The final form of Maxwell-type gravity field equation is therefore given by 
\begin{align}
-\frac{\delta S}{\delta\phi^j}T^{j\mu}_{\ \,\rho}
&= J^\mu_{\,\rho} -\kappa^{-1} \tilde g^{\mu\nu}\partial_\nu b_\rho 
-\partial_\nu\widetilde\calF^{\mu\nu}_{\ \ \rho}\,.
\label{eq:Maxwell-typeQGEq}
\end{align} 
in place of previous classical one (\ref{eq:Maxwell-typeGEq}). 
The expressions Eq.~(\ref{eq:fieldstrength}) 
for the field-strength 
$\widetilde\calF^{\mu\nu}{}_{\!\!\rho}$ and Eq.~(\ref{eq:TildeTk}) for  
the quantities ${\widetilde\calK_{k+1}}^{\soeji\nu_1k \nu\mu}{}_{\!\!\rho}$ 
remain the same as before. Here ${\cal L}$ is understood to be the total Lagrangian 
but actually only the classical Lagrangian part contributes there since all 
the ghost fields $\phi^M=\{b_\mu, \bar c_\mu, c^\mu\}$ have vanishing 
contributions since ${T_M}^\mu_\rho=0$ for them. 
That is, {\it the field-strength 
is in fact the same as that in the classical theory} 
with Lagrangian $L_{\rm cl}$.

\makeatletter
\newcommand{\doublewidetilde}[1]{{%
  \mathpalette\double@widetilde{#1}%
}}
\newcommand{\double@widetilde}[2]{%
  \sbox\z@{$\m@th#1\widetilde{#2}$}%
  \ht\z@=.9\ht\z@ \wd\z@=1.2\wd\z@%
  \widetilde{\box\z@}\!%
}
\makeatother

One may wonder that the final Maxwell-type gravity equation of motion 
(\ref{eq:Maxwell-typeQGEq}) is slightly different from the Yang-Mills case 
since the present ghost field term $\kappa^{-1} \tilde g^{\mu\nu}\partial_\nu b_\rho$ is not 
written in a {\it BRS exact form} like $\{ Q_\B,  D_\mu\bar c\}$ in the latter.
It is actually possible to rewrite Eq.~(\ref{eq:Maxwell-typeQGEq}) into such a 
form. Indeed, the term $\kappa^{-1} \tilde g^{\mu\nu}\partial_\nu b_\rho$ is in fact 
BRS exact up to a divergence of an antisymmetric tensor: 
\begin{equation}
-\kappa^{-1} \tilde g^{\mu\nu}\partial_\nu b_\rho= 
\Brs\bigl(i\kappa^{-1} \tilde g^{\mu\nu}\partial_\nu\bar c_\rho\bigr)
-\partial_\nu\Big(i
(c^\mu\tilde g^{\nu\sigma}-c^\nu\tilde g^{\mu\sigma}) \partial_\sigma\bar c_\rho\Bigr) \ .
\end{equation}
The gravity field equation (\ref{eq:Maxwell-typeQGEq}), therefore,  can be 
rewritten into quite a similar form as the Maxwell-type YM equation:
\begin{equation}
\partial_\nu\doublewidetilde{\calF}{}^{\mu\nu}_{\ \ \rho} 
+ \left\{ \QB, \kappa^{-1} \tilde g^{\mu\nu}\partial_\nu\bar c_\rho\right\}
= J^\mu_{\,\rho} \, ,
\label{eq:MaxwellYMtypeEq}
\end{equation}
where we have written $\Brs( \cdots) =\{ i\QB, \cdots\}$ in terms of the 
BRS charge $\QB$ and defined a modified field strength 
$\doublewidetilde{\calF}{}^{\mu\nu}_{\ \ \rho}$:
\begin{equation}
\doublewidetilde\calF{}^{\mu\nu}_{\ \ \rho}=
\widetilde\calF^{\mu\nu}_{\ \ \rho} + i
(c^\mu\tilde g^{\nu\sigma}-c^\nu\tilde g^{\mu\sigma}) \partial_\sigma\bar c_\rho\,.
\end{equation}
This form of Maxwell-type equation (\ref{eq:MaxwellYMtypeEq}) with the 
BRS exact term was also derived for the Einstein theory case by 
Nakanishi\cite{Nakanishi:1981fj}.

\section{
Noether Current for the Choral symmetries 
in a generic Higher Derivative System}

BRS symmetry, or more generally, choral symmetries $\IOSp(8|8)$ 
exist for any GC 
transformation invariant systems if one adopts the 
gauge-fixing Lagrangian (\ref{eq:NakanishiGF}) of de Donder-Nakanishi gauge\cite{Nakanishi:1990qm}.
This is because the currents of the choral symmetries 
are conserved as far as the equations of motion
\begin{equation}
\partial_\mu(\tilde g^{\mu\nu}\partial_\nu X^M) =0 
\label{eq:16}
\end{equation}
hold for the 16 component `fields' (=4d coordinate $x^\mu$ and three fields)
\cite{Nakanishi:1980rf,Nakanishi:1980db}
\begin{equation}
X^M = ( \hat x^\mu,\ b_\mu,\ c^\mu,\ \bar c_\mu), \qquad \hat x^\mu\equiv x^\mu/\kappa.
\end{equation}
Indeed, this equation of motion for the coordinate $X^M \propto x^\rho$ 
actually implies the de Donder condition on the gravity field:
\begin{equation}
\partial_\mu(\tilde g^{\mu\nu}\partial_\nu x^\rho) =\partial_\mu\tilde g^{\mu\rho}=0 \, .
\end{equation}
And the FP ghost equations of motion 
\begin{equation}
\frac{\partial{\cal L}_\FP}{\partial\bar c_\rho}= -i\partial_\mu(\tilde g^{\mu\nu}\partial_\nu c^\rho) =0, \qquad 
\frac{\partial{\cal L}_\FP}{\partial c^\rho}= +i\partial_\mu(\tilde g^{\mu\nu}\partial_\nu\bar c_\rho) =0
\end{equation}
directly follow from the gauge-fixing Lagrangian (\ref{eq:NakanishiGF}), 
implying the equations for $X^M=c^\rho$ and $\bar c_\rho$. The equation for 
$X^M=b_\rho$ may be a bit non-trivial, but we now already know the 
Maxwell-type 
gravity equation of motion (\ref{eq:Maxwell-typeQGEq}), the divergence 
$\partial_\mu$ of which immediately leads to 
\begin{equation}
\partial_\mu(\tilde g^{\mu\nu}\partial_\nu b_\rho) =0\,.
\end{equation}

These 16 components' {\it d'Alembert's equations of motion} hold 
if and only if the gauge fixing Lagrangian ${\cal L}_\GFP$ is given by the 
de Donder-Nakanishi's one (\ref{eq:NakanishiGF}),
\footnote{
The following discussion on the $\OSp(8|8)$ invariance may be viewed as a 
mere recapitulation of Nakanishi's paper\cite{Nakanishi:1980db,Nakanishi:1990qm}
but we have simplified and made in particular the signs and $i$ factors 
more tractable by introducing a hermitian $\OSp(8|8)$ metric (\ref{eq:OSpmetric}). The derivation of the $\OSp(8|8)$ Noether current in higher derivative 
system is of course new.} 
which can be written in a manifestly $\OSp(8|8)$ invariant form:
\begin{align}
{\cal L}_{\GF+\FP} &= -\kappa^{-1}h E = -\kappa^{-1}h g^{\mu\nu}E_{\mu\nu},  \nn
E_{\mu\nu} 
&= \frac12 \bigl(
\partial_\mu b_\nu+ i\kappa\partial_\mu\bar c_\rho\cdot \partial_\nu c^\rho 
+(\mu\leftrightarrow \nu)\bigr)
\nn
&= \frac12 \bigl(
\partial_\mu b_\rho\cdot \partial_\nu x^\rho+ i\kappa\partial_\mu\bar c_\rho\cdot \partial_\nu c^\rho 
+(\mu\leftrightarrow \nu)\bigr)
\nn
 &= \frac{\kappa}2 
\eta_{\NM}
\partial_\mu X^M \partial_\nu X^N \, .
\label{eq:OSpInvGFP}
\end{align}   
where $\eta_{\MN}$ is the $\OSp(8|8)$ metric given by
\begin{equation}
\eta_{\NM}=
\left(
\begin{array}{cc|cc}
             & \delta_\mu^\nu&    &    \\ 
\delta^\mu_\nu&              &    &    \\ 
\hline
    &        &               & i\delta_\mu^\nu\\ 
    &        & -i\delta^\mu_\nu&                
\end{array}
\right)
=\eta^{\NM} \quad  (\text{inverse})\,.    
\label{eq:OSpmetric}
\end{equation}
Note the symmetry property of this ($c$-number) metric
\begin{equation}
\eta_{\MN}=(-)^{|M|\cdot|N|}\eta_{\NM}= (-)^{|M|}\eta_{\NM}=(-)^{|N|}\eta_{\NM}
\equiv\tilde\eta_{\NM} \, ,
\label{eq:eta-property}
\end{equation}
where the statistics index $|M|$ is 0 or 1 when $X^M$ is bosonic or 
fermionic, respectively. This property (\ref{eq:eta-property}) is 
because $\eta_{\MN}$ is `diagonal' 
in the sense that its off-diagonal, bose-fermi and fermi-bose, 
matrix elements vanish, i.e., $\eta_{\MN}=0$ when $|M|\not=|N|$, so that 
$|M|=|N|=|M|\cdot|N|$ in front of $\eta_{\MN}$.
Note also that the $\tilde\eta_{\NM}$ introduced here is 
just the transposed metric $\eta^\T_{\NM}=\eta_{\MN}$. 
So we have
\begin{eqnarray}
\frac{\kappa}2 \eta_{\NM}X^M X^N 
&=&\frac{\kappa}2 X^M \tilde\eta_{\MN} X^N \nn
&=&\frac{\kappa}2
\left( \hat x^\mu\  b_\mu\ c^\mu\ \bar c_\mu\right)
\left(
\begin{array}{cc|cc}
             & \delta_\mu^\nu&    &    \\ 
\delta^\mu_\nu&              &    &    \\ 
\hline
    &        &               & -i\delta_\mu^\nu\\ 
    &        & +i\delta^\mu_\nu&                
\end{array}
\right)    
\begin{pmatrix}
\hat x^\nu\\
b_\nu\\
c^\nu\\
\bar c_\nu 
\end{pmatrix} \nn
&=& \frac{\kappa}2
\bigl( \hat x^\mu b_\mu+ b_\mu\hat x^\mu-ic^\mu\bar c_\mu+i\bar c_\mu c^\mu\big)
= b_\mu x^\mu+i\kappa\bar c_\mu c^\mu\,.
\nonumber
\end{eqnarray}

Noting the d'Alembert's equations of motion for $X^M$, 
Nakanishi constructed the conserved 
currents\footnote{Our current $\calM^{\MN \mu}$ presented 
here is not exactly equal to Nakanishi's original one\cite{Nakanishi:1990qm} 
$\calM^\mu(X^M, X^N)$, but the precise relation reads 
\begin{equation}
\calM^\mu(X^M, X^N) = i^{|M|\cdot|N|}\calM^{\MN \mu} \, .
\end{equation}
}
\begin{align}
\calM^{\MN \mu} &\equiv\tilde g^{\mu\nu}\bigl ( 
X^M \overset{\leftrightarrow}{\partial}_\nu X^N \bigr) 
\label{eq:OSpNoetherCurrent}
\\
\calP^{M \mu}&\equiv 
\tilde g^{\mu\nu}\partial_\nu X^M =  
\tilde g^{\mu\nu}\bigl( 1 \overset{\leftrightarrow}{\partial}_\nu X^M \bigr) \, . 
\end{align}
He showed from the equal-time commutation relations (ETCR) 
derived in the Einstein gravity theory that their charge 
operators 
\begin{align}
M^{\MN} &\equiv\int d^3x\, \calM^{\MN 0} = (-)^{1+|M|\cdot|N|}M^{\NM}\,, \nn
P^M &\equiv\int d^3x\, \calP^{M 0} 
\end{align} 
generate the following transformations on all the fields 
$\Phi$, gravity and matter fields $\phi^j$ as well as the $\OSp(8|8)$ 
ghost-`fields' 
$X^M = (\hat x^\mu, b_\mu, c^\mu, \bar c_\mu)$: 
\begin{align}
[iM^{\MN}, \Phi\} &\equiv 
\Trs^{\MN}\Phi= \BrsN^{\MN}\Phi-\kappa(\BrsN^{\MN}\!{\hat x^\rho})\partial_\rho\Phi,
\label{eq:ChoralTrf}
\end{align} 
where the Nakanishi transformation $\BrsN^{\MN}$ is an $\OSp$ rotation  
for the $\OSp(8|8)$ ghost-fields $X^L$ given by 
\begin{align}
\BrsN^{\MN}X^L &= -\tilde\eta^{\NL}X^M +(-)^{|M|\cdot|N|}\tilde\eta^{\ML}X^N \,.
\label{eq:OSpTrf}
\end{align}
This transformation, in particular, gives for the coordinate $\hat x^\rho$ 
\begin{align}
\BrsN^{\MN}{\hat x^\rho} 
&= -\tilde\eta^{N\hat x^\rho}X^M +\tilde\eta^{M\hat x^\rho}X^N
\equiv(-\kappa)^{-1}\calE^{\MN\rho} \, ,
\end{align}
which is nonvanishing only when $X^M$ or/and $X^N$ is $b_\mu$. 
And the Nakanishi transformation of the 
gravity/matter fields $\phi^j$ is given by
\begin{align}
\BrsN^{\MN}\phi^j &= \partial_\mu\calE^{\MN\nu}\cdot\left[\phi^j\right]^\mu_{\,\nu} \, .
\end{align}
Therefore, if either $X^M$ or $X^N$ equals $b_\mu$, 
the $\Trs^{\MN}$ transformation 
is just the GC transformation 
with transformation parameter 
$\varepsilon^\rho(x)\rightarrow\calE^{\MN\rho} = -\kappa(\BrsN^{\MN}\!{\hat x^\rho})$ 
for the gravity/matter fields $\phi^j$,
\begin{align}
\Trs^{\MN} \phi^j &= \calE^{\MN\rho}G^j_\rho 
+ \partial_\mu\calE^{\MN\rho}\cdot T^{j\,\mu}_{\ \rho} 
\qquad (G^j_\rho=\partial_\rho\phi^j, \ \ \ T^{j\,\mu}_{\ \ \rho} = \left[\phi^j\right]^\mu_{\,\rho}),   
\nn 
&\text{with field dependent parameter}\ \ 
\calE^{\MN\rho}=-\kappa(\BrsN^{\MN}\!{\hat x^\rho}) \, ,
\label{eq:ChoralTrfonMatter}
\end{align}
and, for the $\OSp(8|8)$ ghost-fields $X^L$, the 
GC transformation as scalar fields {\it plus} an $\OSp$ rotation:
\begin{align}
\Trs^{\MN} X^L &= \calE^{\MN\rho}G^L_\rho 
- (\tilde\eta^{\NL}X^M -(-)^{|M|\cdot|N|}\tilde\eta^{\ML}X^N) 
\qquad (G^L_\rho=\partial_\rho X^L) \, .
\label{eq:ChoralTrfonGhost}
\end{align}
Note here that the `transformation parameter' $\calE^{\MN\rho}$ 
may now be fermionic 
when $|M|+|N|=1$. We have therefore put the factors $G^j_\rho$ and 
$T^{j\mu}_{\ \rho}$ linear in $\phi^j$ behind the parameter 
$\calE^{\MN\rho}$ 
in Eqs.~(\ref{eq:ChoralTrfonMatter}) and (\ref{eq:ChoralTrfonGhost}) 
to avoid the sign factor $(-1)^{|\phi^j|(|M|+|N|)}$. 


The choral invariance of our total Lagrangian ${\cal L}={\cal L}_{\rm cl}+{\cal L}_{\GF+\FP}$ 
is now clear; the gravity/matter fields receives just a special 
GC transformation with parameter $\varepsilon^\rho(x)=\calE^{\MN\rho}$ in 
Eq.~(\ref{eq:ChoralTrfonMatter})
and so the ${\cal L}_{\rm cl}$ part is invariant. 
The gauge fixing Lagrangian ${\cal L}_{\GFP}$ is also clearly invariant 
since in this form of the Lagrangian (\ref{eq:OSpInvGFP}), the $\OSp$-vector 
field components $X^M$ including the co-ordinate $\hat x^\rho$ are 
treated as scalar fields and hence $E_{\mu\nu}$ is clearly $\mu\nu$ tensor 
and $h g^{\mu\nu}E_{\mu\nu}$ is manifestly a GC scalar density, 
and, moreover, $E_{\mu\nu}$ written in the form (\ref{eq:OSpInvGFP}) 
is manifestly invariant under (global) $\OSp(8|8)$ rotation. 
Note that this invariance is made manifest by making the mere parameter 
coordinate $x^\mu$ transforms as if being a field both under the 
GC transformation and the $\OSp$ rotation; actually, those two 
transformations on $\hat x^\mu$ cancels 
each other and the coordinate $x^\mu$ remains intact under $\Trs^{\MN}$ as 
any non-field parameters should be: indeed, Eq.~(\ref{eq:ChoralTrf}) 
indicates for $\Phi=\hat x^\mu$,
\begin{equation}
\Trs^{\MN}{\hat x^\mu}=
\BrsN^{\MN}\!{\hat x^\mu} -\kappa(\BrsN^{\MN}\!{\hat x^\rho})\partial_\rho{\hat x^\mu}=0 \, ,
\end{equation}
where the first term is the $\OSp$ rotation and the second term is 
the GC transformation of $\hat x^\mu$ regarded as a `scalar field'.

Let us now compute the Noether currents corresponding to these choral 
symmetries in our general higher derivative GC invariant system. 
We shall show that the Noether currents coincides with the 
Nakanishi's simple form (\ref{eq:OSpNoetherCurrent}) aside from the 
divergence of an antisymmetric tensor.   

To do this systematically, we devise a local version of the 
choral symmetry transformation (\ref{eq:ChoralTrf}), 
or (\ref{eq:ChoralTrfonMatter}) and (\ref{eq:ChoralTrfonGhost}).  
We multiply them by a local graded  
transformation parameter $\varepsilon_{\NM}(x)$ {\it from the left} 
so that it reduces to the original 
$\OSp(8|8)$ transformation in the global limit 
$\varepsilon_{\NM}(x)\rightarrow\varepsilon_{\NM}\text{:const.}$; namely, we define the 
transformation, 
\begin{equation}
\begin{array}{cll}
\delta\phi^j = 
&\hspace{-.5em}(\varepsilon_{\NM}{\cal E}^{\MN \rho})\,G^j_\rho 
+ \partial_\mu(\varepsilon_{\NM}{\cal E}^{\MN \rho})\cdot T^{j\,\mu}_{\ \rho} 
& \text{for gravity/matter fields} \ \phi^j, \\
\\
\delta X^L = 
&\hspace{-.5em}(\varepsilon_{\NM}{\cal E}^{\MN\rho})\,G^L_\rho 
\\
& - \varepsilon_{\NM}(\tilde\eta^{\NL}X^M 
 -(-)^{|M|\cdot|N|}\tilde\eta^{\ML}X^N)
&\text{for $\OSp$ coordinate fields $X^L$}. 
\end{array}
\label{eq:TrfonGhost}
\end{equation}
We take our parameter $\varepsilon_{\NM}$ Grassmann even or odd according to 
$|N|+|M|= 0$ or 1, respectively, so that 
the product $(\varepsilon_{\NM}{\cal E}^{\MN\rho})$ always becomes an ordinary 
bosonic `parameter' and hence can be moved to anywhere without 
worrying about sign changes. Note, however, that, 
in order to obtain correctly 
the Noether current corresponding to the $\OSp$ transformation 
$\Trs^{\MN}$ in Eq.~(\ref{eq:ChoralTrf}), or 
Eqs.~(\ref{eq:ChoralTrfonMatter}) 
and (\ref{eq:ChoralTrfonGhost}), we have to factor out the 
parameter $\varepsilon_{\NM}$ from the left since it is multiplied from the 
left here. However, the general procedure explained in the 
previous sections to derive Noether current in the higher derivative 
theories, which we follow now, have placed the transformation 
parameter at {\it the most right end}, and the troublesome point is that 
the transformation parameter for the $\OSp$ transformation is the graded 
one $\varepsilon_{\NM}$ but not the bosonic product $(\varepsilon_{\NM}{\cal E}^{\MN\rho})$. 
It is necessary to move those graded quantities separately and freely to 
apply the general procedure to this case, although the transformation 
parameter $\varepsilon_{\NM}$ has eventually to be factored out from the left. 
The best way to forget about the bothering sign factors appearing  
in changing the order of graded quantities, is to adopt a convention 
similar to the so-called 
{\it `implicit grading'}\cite{Butter:2009cp}. 
We take as a {\it natural order} of those graded quantities, 
$\varepsilon_{\NM}$ the first, 
${\cal E}^{\MN\rho}$ second and the other graded quantities like $X^M$, 
$G^J_{\rho}$ and $T^{j\\mu}{}_\rho$ third. Initially, these quantities appear 
in this natural order, since the product factor 
$(\varepsilon_{\NM}{\cal E}^{\MN\rho})$ appearing in Eq.~(\ref{eq:TrfonGhost}) is 
bosonic and can be placed on the most left in any case. Then from this 
natural order we freely move those factors separately anywhere 
without writing any sign factors. Implicit grading scheme means that 
the correct sign factors should be recovered when necessary; that is, 
in any terms containing those graded quantities, the necessary sign factor 
can be found by counting how many times of changing order are necessary 
to bring those factors into the natural order. We adopt hereafter this 
implicit grading scheme. 


We should note that the GC transformation part of 
this transformation (\ref{eq:TrfonGhost}) 
now takes exactly the same form as the GC transformation 
(\ref{eq:gaugetrf}) with (bosonic) transformation parameter
$\varepsilon^\rho(x)\equiv 
\varepsilon_{\NM}{\calE}^{\MN\rho}$:
\begin{equation}
\delta\Phi^j= G^j_\rho\,\varepsilon^\rho+ T^{j\,\mu}_{\ \rho}\partial_\mu\varepsilon_\rho 
\end{equation}
for all the fields $\Phi^j=(\phi^j, X^M)$, and so the total action is still 
invariant, meaning that the total Lagrangian transforms as a scalar 
density: 
$\delta{\cal L}=\partial_\mu({\cal L}\varepsilon^\mu)=\partial_\mu({\cal L}\,\varepsilon_{\NM}{\calE}^{\MN\rho})$. 
 
As for the rest $\OSp(8|8)$ rotation part 
of the $\OSp$ coordinate `fields' $X^L$, 
\begin{equation}
\hat \delta^{\OSp} X^L \equiv 
-\varepsilon_\NM \bigl( \tilde\eta^{\NL}X^M 
 -(-)^{|M|\cdot|N|}\tilde\eta^{\ML}X^N \bigr) \, ,
\label{eq:OSpRot}
\end{equation}
however, the ${\cal L}_\GFP$ in 
Eq.~(\ref{eq:OSpInvGFP}) is no longer invariant under the rotation with 
{\it $x$-dependent} parameter $\varepsilon_\MN$. As shown explicitly in the 
Appendix A, we can 
immediately find the change of ${\cal L}_\GFP$ as 
Eq.~(\ref{eq:OSpChangeOfLGFP}):
\begin{equation}
\delta{\cal L}_\GFP = 
{\tilde g^{\mu\nu}}\,\partial_\mu\varepsilon_\NM \cdot 
\bigl( X^M\overset{\leftrightarrow}{\partial}_\nu X^N \bigr)
={\tilde g^{\mu\nu}}
\bigl( X^M\overset{\leftrightarrow}{\partial}_\nu X^N \bigr)
\,\partial_\mu\varepsilon_\NM \, , 
\end{equation}
where note that we have already used `implicit grading' at the last 
equality. Following this implicit grading scheme, 
we can write the change of our total Lagrangian 
${\cal L}={\cal L}_{\rm cl}+{\cal L}_\GFP$ 
under our transformation (\ref{eq:TrfonGhost}) in the form 
\begin{align}
\delta{\cal L} &= \partial_\mu({\cal L}\,{\calE}^{\MN\mu}\varepsilon_{\NM}) 
+{\tilde g^{\mu\nu}}
\bigl( X^M\overset{\leftrightarrow}{\partial}_\nu X^N \bigr) 
\cdot \partial_\mu\varepsilon_\NM \nn
&= \partial_\mu({\cal L}\,\calE^{\MN\mu})\cdot\varepsilon_\NM + 
\Bigl({\cal L}\,\calE^{\MN\mu}+{\tilde g^{\mu\nu}}
\bigl( X^M\overset{\leftrightarrow}{\partial}_\nu X^N \bigr)
\Bigr) \partial_\mu\varepsilon_\NM \ . 
\label{eq:LchangeUnderBRS}
\end{align}

Now we can rewrite our transformation in 
the same form as the general gauge transformation (\ref{eq:gaugetrf}), 
which contains the zero-th and first order differentiation 
of the transformation parameter $\varepsilon_\NM$; that is, it is 
unifiedly given 
for gravity/matter and ghost fields $\Phi^I=( \phi^j, X^L )$ 
in the form
\begin{equation}
\delta\Phi^I = \calG^{I\MN}\varepsilon_{\NM}+\calT^{I\MN\mu}\partial_\mu\varepsilon_{\NM} \, ,
\label{eq:BRS-Trf} 
\end{equation}
where the coefficients $\calG^{I\MN}$ and $\calT^{I\MN\mu}$ are 
given as
\begin{align}
\calG^{I\MN}&= 
\begin{cases}
G^j_\rho{\calE}^{\MN \rho}
+T^{j\,\mu}_{\ \ \rho}\partial_\mu{\calE}^{\MN \rho} 
& \text{for $\Phi^I=\phi^j$: gravity/matter} \\
G^L_\rho\,{\calE}^{\MN\rho} & \\
\ {}- \bigl( \tilde\eta^{\NL}X^M 
 -(-)^{|M|\cdot|N|}\tilde\eta^{\ML}X^N \bigr)
& \text{for $\Phi^I=X^L$: ghost}
\end{cases}
\nn
\calT^{I\MN\mu}&= 
\begin{cases}
T^{j\,\mu}_{\ \ \rho}{\calE}^{\MN \rho}
& \text{for $\Phi^I=\phi^j$: gravity/matter} \\
0
& \text{for $\Phi^I=X^L$: ghost}
\end{cases}.
\label{eq:BRS-GT}
\end{align}
So we can now follow the general discussions presented 
in the previous two sections to derive the Noether currents in 
this system.  We should also note that 
the Lagrangian change $\delta{\cal L}$ is now given 
by Eq.~(\ref{eq:LchangeUnderBRS}) in place of Eq.~(\ref{eq:ChangeL}). 
Then, we see that the previous $\delta{\cal L}$-identities following from 
the coefficients of $n$-th order 
derivatives $\partial_{\soeji\mu_1n}\varepsilon^\rho$ of the transformation 
parameter $\varepsilon^\rho$,
Eqs.~(\ref{eq:L0}) and (\ref{eq:L2}) now also hold 
with understanding that 
the coefficients $G_\rho$ and $T^\mu_\rho$ of the gauge transformation 
(\ref{eq:gaugetrf}) there are now replaced by the present transformation's 
ones: That is, by making the replacement 
\begin{equation}
G^j_\rho\ \rightarrow\  \calG^{I\MN}, \qquad T^{j\mu}_\rho\ \rightarrow\ \calT^{I\MN\mu}, 
\end{equation}
and taking account of the form of $\delta{\cal L}$ given 
in Eq.~(\ref{eq:LchangeUnderBRS}), 
we find the identities as coefficients of $n$-th order derivatives 
$\partial_{\soeji\mu_1n}\varepsilon_\NM$ with $n=0, 1$ and $n\geq2$, 
respectively, 
\begin{align}
&\sum_{n=0}^N {\cal L}_I^{\,;\soeji\mu_1n}\partial_{\soeji\mu_1n}\calG^{I\MN}
-\partial_\rho({\cal L}\,\calE^{\MN\rho}) =0 
\label{eq:BRSL0}\\
&{\calK_0}^{\MN\mu} 
-{\cal L}\,\calE^{\MN\mu}
-{\tilde g^{\mu\nu}}
\bigl( X^M\overset{\leftrightarrow}{\partial}_\nu X^N \bigr)=0
\label{eq:BRSL1}\\
& 
\mathop{\text{Sym}}_{\{\soeji\nu_1k \mu\}}
\Bigl({\calK_k}^{\MN\soeji\nu_1k \mu}\Bigr) =0 \qquad \text{for}\quad 
k=1,2,\cdots \ ,
\label{eq:BRSL2}
\end{align}
where 
${\calK_k}^{\MN\soeji\nu_1k\mu}$ now reads
\begin{align}
{\calK_k}^{\MN\soeji\nu_1k \mu}
\equiv 
&\sum_{n=0}^{N-k-1}
{}_{n+k+1}C_{k+1}\, {\cal L}_I^{\,;\soeji\nu_1k \mu\soeji\alpha_1n}\,\partial_{\soeji\alpha_1n}
\calG^{I\MN}
\nn 
&+\sum_{n=0}^{N-k}
{}_{n+k}C_{k}\, {\cal L}_I^{\,;\soeji\nu_1k\soeji\alpha_1n}\,\partial_{\soeji\alpha_1n}
\calT^{I\MN\mu} \ .
\label{eq:defCalK}
\end{align}
Then, by combining equation of motion (\ref{eq:DerS}), we firstly obtain 
a conservation equation for the $\OSp(8|8)$ Noether current from the 
0-th order $\delta{\cal L}$-identity (\ref{eq:BRSL0}) as an analogue of 
Eq.~(\ref{eq:NoetherCurrent}), 
or (\ref{eq:GCNoetherCurrentQuantum}) 
\begin{align}
-{\delta S\over\delta\Phi^I}\calG^{I\MN} &= \partial_\mu J^{\MN\mu} \nn
J^{\MN\mu} &= \sum _{n=0}^{N-1}
{\cal L}_I^{\,;\mu\soeji\alpha_1n}
\bothDer{}{\longLRarrow}{\soeji\alpha_1n}\calG^{I\MN} 
- {\cal L}\,\calE^{\MN\mu} \, .
\label{eq:BRSNoetherCurrent}
\end{align}
Secondly, as an analogue of Eq.~(\ref{eq:Maxwell-typeGEq}), or 
Eq.~(\ref{eq:Maxwell-typeQGEq}), we find the desired 
equation from the first order $\delta{\cal L}$-identity (\ref{eq:BRSL1}) 
and second or higher order $\delta{\cal L}$-identity (\ref{eq:BRSL2}):
\begin{align}
-\frac{\delta S}{\delta\Phi^I}\calT^{I\MN\mu}
&= J^{\MN\mu} 
-{\tilde g^{\mu\nu}}
\bigl(  X^M\overset{\leftrightarrow}{\partial}_\nu X^N \bigr)
- \partial_\nu\widetilde\calF^{\MN\mu\nu}.
\label{eq:Maxwell-typeBRSCurrentEq}
\end{align} 
where $\widetilde\calF^{\MN\nu\mu}$ is the antisymmetric tensor 
(as an ambiguity term of $\OSp$ Noether current) given by 
\begin{equation}
\widetilde\calF^{\MN\nu\mu}
=
\sum_{k=0}^{N-1} (-)^k
\partial_{\soeji\nu_1k}{\widetilde\calK_{k+1}}^{\MN\soeji\nu_1k\nu\mu}
\label{eq:BRSAntisymmetricTensor}
\end{equation} 
with 
\begin{align}
&\hspace{-1.2em}{\widetilde\calK_{k+1}}^{\MN\soeji\nu_1k \nu\mu} \nn
&\equiv 
\frac{k+1}{k+2}
\sum_{n=0}^{N-k-1}
{}_{n+k+1}C_{k+1}\, 
\left(
{\cal L}_I^{\,;\soeji\nu_1k\nu\soeji\alpha_1n}\,\partial_{\soeji\alpha_1n}\calT^{I\MN\mu} 
- (\nu\leftrightarrow \mu)
\right) \nn
&\text{for}\quad k=0, 1, 2, \cdots\ .
\label{eq:BRSTildeTk}
\end{align}

The Eq.~(\ref{eq:Maxwell-typeBRSCurrentEq}) shows that the 
Noether current of $\OSp(8|8)$ symmetry takes the form on-shell,
\begin{equation}
J^{\MN\mu}= 
{\tilde g^{\mu\nu}}
\bigl( X^M\overset{\leftrightarrow}{\partial}_\nu X^N \bigr) 
+\partial_\nu\widetilde\calF^{\MN\mu\nu} \ ,
\end{equation}
so that the charge can be given by the Nakanishi's simple form
\begin{equation}
M^\MN = \int d^3x 
\ h \correct{g^{0\mu}}{g^{0\nu}}
\bigl( X^M\overset{\leftrightarrow}{\partial}_\nu X^N \bigr)
\label{eq:OSpcharge} 
\end{equation}
as symmetry generators for any local operators.%
\footnote{It is, however, quite another 
problem whether the RHS volume integral of (\ref{eq:OSpcharge}) gives a 
well-defined charge even when the symmetry is spontaneously unbroken. 
Generally, the volume integral converges only when the current contains the 
ambiguity term $\partial_\nu\widetilde\calF^{\MN\mu\nu}$ with a suitable coefficient, 
which is in fact a key point when discussing the spontaneous breaking or 
non-breaking of the symmetry of the charge.} 

Finally, recall that we are adopting implicit grading. Then, 
the factor $\calE^{\MN\mu}$ or its derivatives contained in 
$\calG^{I\MN}$ or $\calT^{I\MN\mu}$ in the definitions of the 
current $J^{\NM\mu}$, (\ref{eq:BRSNoetherCurrent}), the field strength 
$\tilde\calF^{\MN\nu\mu}$, (\ref{eq:BRSAntisymmetricTensor}), 
and $\tilde\calK_{k+1}$, (\ref{eq:BRSTildeTk}), should be placed at the 
most left, and if they are kept at the place as written, then the 
grading sign factor should be put necessary for bringing them there 
from the most left. Fortunately, however, 
the sign factors actually turn out to be unnecessary here.  
This is because the other graded quantities to jump over when bringing 
the factor $\calE^{\MN\mu}$ to the most left are essentially only 
the bosonic Lagrangian ${\cal L}$ as a whole. For instance, 
the Noether current (\ref{eq:BRSNoetherCurrent}) contains the terms 
\begin{equation}
{\cal L}_I^{\,;\mu\soeji\alpha_1n}
\bothDer{}{\longLRarrow}{\soeji\alpha_1n}\calG^{I\MN} 
- {\cal L}\,\calE^{\MN\mu} .
\end{equation}
In the second term, it is the Lagrangian ${\cal L}$ itself for 
$\calE^{\MN\mu}$ to jump over. In the first term, on the other hand, 
${\cal L}_I^{\,;\mu\soeji\alpha_1n}=\partial{\cal L}/\partial_{\mu\soeji\alpha_1n}\phi^I$ is fermionic when 
the field $\phi^I$ is a fermion. But, it is immediately followed by the 
factor $G_\rho^I=\partial_\rho\phi^I$ in the 
$\calG^{I\MN}=G_\rho^I \calE^{\MN\rho}$, so that the net factor 
${\cal L}_I^{\,;\mu\soeji\alpha_1n}G_\rho^I$ in front of $\calE^{\MN\rho}$ is 
always bosonic, carrying the same statistics as the original ${\cal L}$.

\section{Conclusion}

In this paper we considered the general GC invariant theory which contains
arbitrarily high order derivative fields. We identified there 
the explicit expression for the energy-momentum as the Noether current 
corresponding to the rigid case of GC transformation, and have shown that 
a linear combination of the equations of motion can be rewritten into 
the form of Maxwell-type field equation which has the total 
energy-momentum as 
its source. This was done both for gauge unfixed classical system and 
for the gauge-fixed quantum system in the de Donder-Nakanishi gauge. 
The Maxwell-type field equation in the latter has formally an additional 
term coming from the gauge-fixing compared with that in the former, which 
was shown to take a BRS-exact form just like in the well-known Yang-Mills case.

By using the same technique, we derived similar expressions of Maxwell-type 
equations for the Noether currents for the $\IOSp(8|8)$ choral symmetries. 
It confirmed that the Nakanishi's original result persists to any GC invariant 
system. 

These results will be useful for proving the existence 
theorem \erase{[?]}\add{\cite{Kugo:1985jc}}
of graviton 
(photon) in any GC (local $U(1)$) gauge invariant system as far as the 
translation (global $U(1)$) symmetry is not spontaneously broken. We hope 
we can report on this matter near future. 

The techniques used for the higher derivative systems in this paper will also 
be useful for studying more general problems. For instance, the basic 
problems such 
as the (non-)equivalence between canonical quantization and path-integral 
quantization 
may be discussed in a general fashion for the higher derivative systems. 
This is so since the present technique can easily be combined with 
Ostrogradsky's canonical formalism for higher derivative 
systems\cite{Ostrogradsky}.

\section*{Acknowledgment}

The author would like to thank Ichiro Oda for valuable discussions on 
Weyl invariant gravity theories and reading the early version of this 
manuscript. This work is supported in part by the JSPS KAKENHI 
Grant Number JP18K03659.

\appendix

\section{$\OSp$ transformation of ${\cal L}_{\GFP}$}
\def\T{\text{T}}

Note that 
the parameter $\varepsilon_{\NM}$ and the metric $\eta_\NM=\eta^\NM$ have 
the following statistics and transposition properties: 
\begin{align}
|\varepsilon_{\NM}| &= |M|+|N|, \qquad \text{i.e.,}\quad 
\varepsilon_{\NM}X^L = (-)^{|L|(|M|+|N|)}X^L\varepsilon_{\NM} \nn
\varepsilon_{\NM}&= (\varepsilon^\T)_{\MN}=(-)^{1+|M|\cdot|N|}\varepsilon_{\MN} \nn
\eta_{\NM}&= (\eta^\T)_{\MN}
=(-)^{|M|}\eta_{\MN} =(-)^{|N|}\eta_{\MN} \equiv\tilde\eta_{\MN}. 
\end{align}

Using these properties and also noting that 
the metric $\eta_\MN$ is a usual c-number (i.e., bosonic) quantity, 
we can rewrite the two terms in the $\OSp$ transformation 
(\ref{eq:OSpRot}) of the $\OSp$ coordinates $X^L$ as 
\begin{align}
-\varepsilon_{\NM} \tilde\eta^{\NL}X^M 
&= -\eta^{\LN}\varepsilon_\NM X^M = - (\eta^{-1} \varepsilon X )^L \nn
+\varepsilon_{\NM}(-)^{|M|\cdot|N|}\tilde\eta^{\ML}X^N
& 
=(-)^{|M|\cdot|N|}\eta^{\LM}(-)^{1+|M|\cdot|N|}\varepsilon_{\MN} X^N \nn
&=-\eta^{\LM}\varepsilon_{\MN} X^N . 
= -(\eta^{-1} \varepsilon X )^L,  
\end{align}
where $X$, $\eta^{-1}$ and $\varepsilon$ are 
regarded as a column vector and matrices in the final expressions. 
Thus the two terms reduce to the same expression and so the 
$\OSp$ transformation (\ref{eq:OSpRot}) is rewritten concisely into
\begin{equation}
\delta^\OSp X^M  = -2(\eta^{-1}\varepsilon X )^M = -2\eta^{\MN}\varepsilon_{\NL} X^L . 
\end{equation}
Now it is easy to see how the gauge-fixing Lagrangian 
${\cal L}_\GFP=-\kappa^{-1}\tilde g^{\mu\nu}E_{\mu\nu}$ changes under the 
$\OSp$ transformation (\ref{eq:OSpRot}).
Noting the expression of $E_{\mu\nu}$ in Eq.~(\ref{eq:OSpInvGFP}), 
we find 
\begin{align}
\delta^{\OSp}{\cal L}_{\GFP}&=\delta^{\OSp}(-\kappa^{-1}\tilde g^{\mu\nu}E_{\mu\nu}) \nn
&= -\frac12\tilde g^{\mu\nu}\eta_{\NM} \delta^{\OSp}(\partial_\mu X^M \partial_\nu X^N) 
= -\tilde g^{\mu\nu}\eta_{\NM} (\delta^{\OSp}\partial_\mu X^M) \partial_\nu X^N \nn
&= -\tilde g^{\mu\nu}\eta_{\NM} \partial_\mu(-2\eta^{-1}\varepsilon X)^M \partial_\nu X^N 
= 2\tilde g^{\mu\nu} \partial_\mu(\varepsilon_{\NM}X^M) \partial_\nu X^N \nn
&= 2\tilde g^{\mu\nu} \Bigl(\partial_\mu\varepsilon_{\NM}\cdot X^M\partial_\nu X^N
+\varepsilon_{\NM}\partial_\mu X^M\partial_\nu X^N \Bigr).
%
\end{align}   
Now we note that the $\OSp$ transformation parameter $\varepsilon_\NM$ must be 
graded antisymmetric 
\begin{equation}
\varepsilon_\NM = - (-)^{|N|\cdot|M|}\varepsilon_\MN
\label{eq:GradedAntisymmetry}
\end{equation}
in order for the Lagrangian to be global $\OSp$ invariant. Indeed, 
only the second term remains in the global transformation, and it 
vanishes if and only if $\varepsilon_\NM$ is graded antisymmetric since 
$\partial_\mu X^M\partial_\nu X^N$, multiplied by $\tilde g^{\mu\nu}$, is graded symmetric.
The second term thus vanishes also here, and the remaining first term 
is rewritten into
\begin{align}
\delta^{\OSp}{\cal L}_{\GFP}
&=2{\tilde g^{\mu\nu}}
\partial_\mu\varepsilon_\NM\cdot\bigl( X^M\partial_\nu X^N \bigr) \nn 
&={\tilde g^{\mu\nu}} \partial_\mu\varepsilon_\NM\cdot\bigl( X^M\partial_\nu X^N 
-(-)^{|M|\cdot|N|}X^N\partial_\nu X^M\bigr) \nn 
&={\tilde g^{\mu\nu}}\partial_\mu\varepsilon_\NM\cdot\bigl( X^M\partial_\nu X^N 
-\partial_\nu X^M\cdot X^N\bigr) \nn 
&={\tilde g^{\mu\nu}}\partial_\mu\varepsilon_\NM\cdot
\bigl( X^M\overset{\leftrightarrow}{\partial}_\nu X^N \bigr), 
\label{eq:OSpChangeOfLGFP}
\end{align}
where, in going to the second line, we have exchanged the dummy 
indices $N\leftrightarrow M$ and used the graded antisymmetry property 
of $\varepsilon_\NM$, (\ref{eq:GradedAntisymmetry}).

\section{$\OSp(2|2)$-invariant scalar field system}

To see the property of the $\OSp$ symmetry, let us consider here the 
$\OSp$-invariant system on flat Minkowski background in 
which scalar fields belongs to an 
$\OSp(2|2)$ vector representation. 
\begin{equation}
{\cal L}= 
-\frac12 \eta_{\NM} \partial_\mu\phi^M \partial^\mu\phi^N,
\label{eq:OSpL}
\end{equation}
where $\phi^M$ is a $2+2$-component $\OSp$-vector whose first 2-components 
are bosons and the rest 2-components are fermions and 
the $\OSp(2|2)$ metric is given by
\begin{equation}
\eta_{\NM}=
\begin{pmatrix}
\sigma_1 &  \\
  & -\sigma_2
\end{pmatrix}
=\eta^{\NM}.
\end{equation}
Note that we have put an overall minus sign to our 
Lagrangian (\ref{eq:OSpL}) in order to make it coincide with the 
convention of the gauge fixing Lagrangian ${\cal L}_\GFP$ in 
Eq.~(\ref{eq:OSpInvGFP}),  
although it is not physically important  in any case
since the $\OSp$ metric 
is neither positive- nor negative- definite.

The infinitesimal $\OSp$ rotation is parametrized as
\begin{equation}
\delta\phi^M = \varepsilon^M_{\ \,N} \phi^N.
\end{equation}
Under this rotation, the quadratic kinetic Lagrangian (\ref{eq:OSpL}) 
is transformed as
\begin{align}
\delta{\cal L}
&= -\frac12 
\eta_{\NM} \bigl(\varepsilon^M_{\ \,L}\partial_\mu\phi^L \partial^\mu\phi^N 
+ \partial_\mu\phi^M \varepsilon^N_{\ \,L}\partial^\mu\phi^L \bigr) \nn
&=-\frac12 \bigl(\eta_{\NM}\varepsilon^M_{\ \,L}\partial_\mu\phi^L \partial^\mu\phi^N 
+ \partial_\mu\phi^M \eta_{\NM}\varepsilon^N_{\ \,L}\partial^\mu\phi^L \bigr) \nn
&=-\frac12 \bigl(\varepsilon_{\NL}\partial_\mu\phi^L \partial^\mu\phi^N 
+ \partial_\mu\phi^M (-)^{|M|}\varepsilon_{\ML}\partial^\mu\phi^L \bigr) \nn
&=-\frac12 \bigl(\varepsilon_{\NL}\partial_\mu\phi^L \partial^\mu\phi^N 
+ \varepsilon_{\ML}\partial^\mu\phi^L\partial_\mu\phi^M  \bigr)
=-\varepsilon_{\NM}\partial_\mu\phi^M \partial^\mu\phi^N 
\end{align}
with $\varepsilon_{\NM}\equiv\eta_{\NL}\varepsilon^L_{\ M}$. This can further be rewritten as
\begin{align}
\varepsilon_{\NM}\partial_\mu\phi^M \partial^\mu\phi^N 
&=\varepsilon_{\MN}\partial_\mu\phi^N \partial^\mu\phi^M 
=(-)^{|M|\cdot|N|}\varepsilon_{\MN}\partial_\mu\phi^M \partial^\mu\phi^N \nn 
&=\frac12 \bigl(
(-)^{|M|\cdot|N|}\varepsilon_{\MN}+ \varepsilon_{\NM}\bigr)\partial_\mu\phi^M \partial^\mu\phi^N \ . 
\end{align}
So, if the transformation parameter $\varepsilon_{\NM}$ is 
graded anti-symmetric, i.e.,  
\begin{equation}
\varepsilon_{\NM} = - (-)^{|M|\cdot|N|}\varepsilon_{\MN},
\end{equation}
then the Lagrangian (\ref{eq:OSpL}) is $\OSp$-invariant. 

We define the canonical conjugate variable $\pi_M$ by the 
right-derivative
\begin{equation}
\pi_M\equiv\partial{\cal L}/\partial\dot\phi^M = -\eta_{\MN}\dot\phi^N \, .
\end{equation} 
Then the ETCR is given by\footnote{If the conjugate momentum $\pi_M$ were 
defined by the left derivative, then the ETCR should be given by
\begin{equation}
[\pi_M, \phi^N\}= -i\delta_M^N \, .
\end{equation}
}

\begin{equation}
[\phi^M, \pi_N\} = i\delta^M_N \ \rightarrow\ 
[\phi^M, \dot\phi^N\} = 
-\eta^{\NL}[\phi^M, \pi_L\} = 
-\eta^{\NL}i\delta^M_L = -i\eta^{\NM} \, ,
\end{equation}
so that
\begin{equation}
[\phi^M, \dot\phi^N\} = -i (-)^N\eta^{\MN} \equiv-i\tilde\eta^{\MN}, \qquad  
[\dot\phi^M, \phi^N\} = i (-)^N\eta^{\MN}= i\tilde\eta^{\MN} \, .  
\end{equation}
The Noether current $J^{\MN\mu}$ for the $\OSp(2|2)$ transformation is 
defined by
\begin{align}
-\frac12 \varepsilon_{\NM}J^{\MN\mu} &\equiv 
\bigl(\partial{\cal L}/\partial(\partial_\mu\phi^M)\bigr)\delta\phi^M
=-\eta_{\MN}\partial^\mu\phi^N \varepsilon^M_{\ \,L}\phi^L \nn
&=-(-)^N\partial^\mu\phi^N \varepsilon_{\NL}\phi^L 
=-\varepsilon_{\NL}\phi^L\partial^\mu\phi^N  \nn
&=-\varepsilon_{\NM}\frac12(\phi^M\partial^\mu\phi^N -(-)^{|N|\cdot|M|} \phi^N\partial^\mu\phi^M) \nn
&=-\varepsilon_{\NM}\frac12(\phi^M\partial^\mu\phi^N - \partial^\mu\phi^M\cdot\phi^N) 
=-\frac12\varepsilon_{\NM}(\phi^M \overset{\leftrightarrow}{\partial^\mu}\phi^N ) \ ,
\end{align}
where we have used the graded-antisymmetry property of $\varepsilon_{\NM}$ 
in going from the second to third lines.  
Thus we have
\begin{equation}
J^{\MN\mu}= \phi^M \overset{\leftrightarrow}{\partial^\mu}\phi^N \, ,
\end{equation}
so that the charge is given by
\begin{equation}
M^{\MN}= \int d^3x J^{\MN 0}
= \int d^3x (\phi^M \dot\phi^N - \dot\phi^M \phi^N) \, .
\end{equation}
This actually generates the original $\OSp$ rotation by ETCR's:
\begin{align}
[iM^{\MN}, \phi^L\} 
&= \int d^3x \bigl( i\phi^M [\dot\phi^N, \phi^L\} 
- i(-)^{|N|\cdot|L|}[\dot\phi^M, \phi^L\} \phi^N \bigr) \nn
&= -\phi^M \tilde\eta^{\NL}+ (-)^{|N|\cdot|L|}\tilde\eta^{\ML}\phi^N \nn
&= -\phi^M \tilde\eta^{\NL}+ (-)^{|N|\cdot|M|}\phi^N\tilde\eta^{\ML} \ , 
\\
{}[iM^{\MN}, \dot\phi^L\} 
&= \int d^3x \bigl( i(-)^{|N|\cdot|L|}[\phi^M, \dot\phi^L\}\dot\phi^N 
- i\dot\phi^M [\phi^N, \dot\phi^L\} \bigr) \nn
&= (-)^{|N|\cdot|L|}\tilde\eta^{\ML}\dot\phi^N 
-\dot\phi^M \tilde\eta^{\NL} \nn
&= -\dot\phi^M \tilde\eta^{\NL} + (-)^{|N|\cdot|M|}\dot\phi^N\tilde\eta^{\ML} \ .
\end{align}
The $\IOSp(2|2)$ algebra is confirmed as
\begin{align}
[iM^{\MN}, M^{\RS}\} 
&= [iM^{\MN}, \int d^3x (\phi^R \dot\phi^S - \dot\phi^R \phi^S) \} \nn
&= \int d^3x \Bigl[
-\phi^M\tilde\eta^{\NR} \dot\phi^S 
+\dot\phi^M\tilde\eta^{\NR} \phi^S  \nn
&+(-)^{(|M|+|N|)|R|}\bigl(
-\phi^R \dot\phi^M\tilde\eta^{\NS}
+\dot\phi^R \phi^M\tilde\eta^{\NS}\bigr) 
- (-)^{|M|\cdot|N|}\Bigl(M\leftrightarrow N\Bigr) \Bigr] \nn
&= 
-M^{\MS}\tilde\eta^{\NR} +(-)^{|R|\cdot|S|} M^{\MR}\tilde\eta^{\NS} \nn
 &\hspace{2em}{} - (-)^{|M|\cdot|N|}\bigl(
-M^{\NS}\tilde\eta^{\MR} +(-)^{|R|\cdot|S|} M^{\NR}\tilde\eta^{\MS}\bigr)\nn
&= 
-M^{\MS}\tilde\eta^{\NR}+ \Bigl(
 \hbox{3 graded anti-symmetrization terms} \nn
 &\hspace{9em}\hbox{under
 $(M\leftrightarrow N)$ and $(R\leftrightarrow S)$}
\Bigr) \ ,
\nn
%
{}[iM^{\MN}, P^R \} 
&= [iM^{\MN}, \int d^3x (\dot\phi^R) \} \nn
&= \int d^3x \Bigl[
-\dot\phi^M\tilde\eta^{\NR} + (-)^{|M|\cdot|N|}\Bigl(M\leftrightarrow N\Bigr)
\Bigr] \nn
&= 
-P^M \tilde\eta^{\NR} + (-)^{|M|\cdot|N|}P^N \tilde\eta^{\MR} \ .
\end{align}

\end{document}